\documentclass[aps,prd,nofootinbib,twocolumn,superscriptaddress,preprintnumbers,groupedaddress,showpacs]{revtex4-1}

\usepackage{amssymb}
\usepackage{amsmath}
\usepackage{epsfig}
\usepackage{hyperref}
\usepackage{comment}
\usepackage{color}
\usepackage{cleveref}
\usepackage{multirow}
\usepackage{capt-of}

\usepackage{comment}
\excludecomment{toexclude}
\includecomment{toinclude}

\usepackage{graphicx}
\usepackage{gensymb}

\usepackage[utf8]{inputenc} 

\usepackage{undertilde}
\makeatletter
\def\simgt{\mathrel{\lower2.5pt\vbox{\lineskip=0pt\baselineskip=0pt
           \hbox{$>$}\hbox{$\sim$}}}}
\def\simlt{\mathrel{\lower2.5pt\vbox{\lineskip=0pt\baselineskip=0pt
           \hbox{$<$}\hbox{$\sim$}}}}
\makeatother

\newcommand{\be}{\begin{equation}}
\newcommand{\ee}{\end{equation}}
\newcommand{\bea}{\begin{eqnarray}}
\newcommand{\eea}{\end{eqnarray}}

\newcommand{\aap}{{Astron.~Astrophys.}}

\definecolor{seagreen}{rgb}{0.180392,0.545098,0.341176}

\begin{document}

\title{Model-Independent Indirect Detection Constraints on Hidden Sector Dark Matter}

\author{Gilly Elor}
\author{Nicholas L. Rodd}
\author{Tracy R. Slatyer}
\author{Wei Xue}
\affiliation{Center for Theoretical Physics, Massachusetts Institute of Technology, Cambridge, MA}

\begin{abstract}
If dark matter inhabits an expanded ``hidden sector'', annihilations may proceed through sequential decays or multi-body final states. We map out the potential signals and current constraints on such a framework in indirect searches, using a model-independent setup based on multi-step hierarchical cascade decays. 
While remaining agnostic to the details of the hidden sector model, our framework captures the generic broadening of the spectrum of secondary particles (photons, neutrinos, $e^+e^-$ and $\bar{p} p$) relative to the case of direct annihilation to Standard Model particles. We explore how indirect constraints on dark matter annihilation limit the parameter space for such cascade/multi-particle decays. We investigate limits from the cosmic microwave background by {\it Planck}, the {\it Fermi} measurement of photons from the dwarf galaxies, and positron data from AMS-02. The presence of a hidden sector can change the constraints on the dark matter by up to an order of magnitude in either direction (although the effect can be much smaller). We find that generally the bound from the {\it Fermi} dwarfs is most constraining for annihilations to photon-rich final states, while AMS-02 is most constraining for electron and muon final states; however in certain instances the CMB bounds overtake both, due to their approximate independence on the details of the hidden sector cascade. We provide the full set of cascade spectra considered here as publicly available code with examples at http://web.mit.edu/lns/research/CascadeSpectra.html.
\end{abstract}


\maketitle

\section{Introduction}

\begin{figure*}[t!]
\centering
  \includegraphics[scale=0.43]{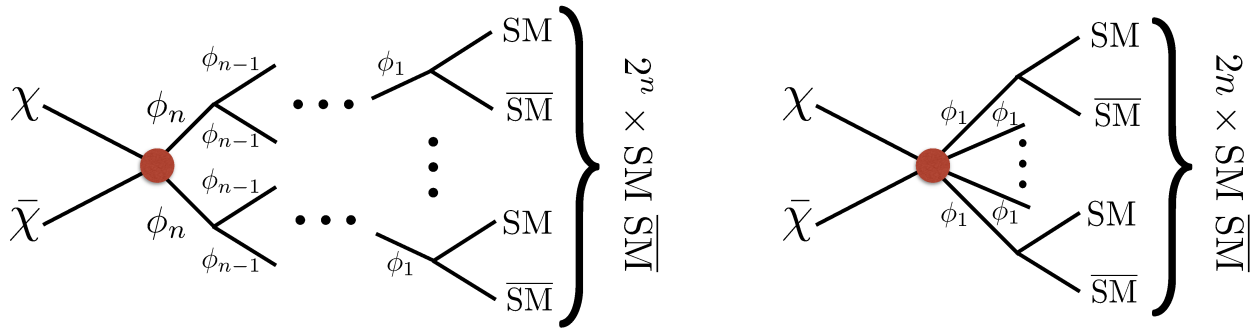}
  \caption{\footnotesize{Left: Schematic diagram of a generic hidden sector cascade. The DM, secluded in its own hidden sector, first annihilates to a pair of hidden sector particles. These $\phi_n$ mediators subsequently decay to lighter particles in the hidden sector and finally to SM particles. Here we consider SM = $\{\gamma, e, \mu, \tau, b, H, W, g\}$ and $n =$ 0-6 step cascades (where $n =$ 0 refers to the usual case of direct annihilations). Right: An equivalent diagram depicting the case where the DM annihilates through an off-shell heavy mediator; effectively decaying to an $n$-body state in the hidden sector which then decays to SM particles.}}
  \label{fig:Cartoon}
\end{figure*}

Indirect searches provide one of the best ways to probe the nature of dark matter (DM) beyond gravitational interactions. Through the observation of gamma rays, cosmic rays, and the anisotropies of the Cosmic Microwave Background (CMB), we may find a hint of DM annihilations to Standard Model (SM) particles. Many models have been proposed in which DM annihilates directly to a pair of SM particles through, for example, a Higgs~\cite{Patt:2006fw,MarchRussell:2008yu}, gauge boson~\cite{Dienes:1996zr}, axion~\cite{Nomura:2008ru}, or neutrino~\cite{Falkowski:2009yz}. Going beyond these simple models, we can consider scenarios in which DM is secluded in its own rich dark sector; such a setup is well motivated from top-down considerations (e.g. \cite{Essig:2013lka} and references therein). In such scenarios, the DM does not couple directly to SM particles (or such couplings are highly suppressed), but instead annihilates to unstable dark sector particles. These states may decay to SM particles or to other dark sector states, but eventually mediator particles that couple to the SM are produced. The mediators subsequently decay into SM particles, which in turn decay to stable and detectable photons, neutrinos, electrons, positrons, protons and/or antiprotons. We refer to this pattern as a ``cascade annihilation'' or simply ``cascade'', with a number of steps given by the number of distinct on-shell dark-sector states between the initial DM annihilation and the production of SM particles. We illustrate this setup schematically in Fig.~\ref{fig:Cartoon}.  

Hidden Sector DM scenarios encompass a broad class of models. For instance models containing one light dark photon mediator \cite{Holdom:1985ag,ArkaniHamed:2008qn,Pospelov:2007mp}, generically give rise to one-step cascades decays; DM annihilates to two dark photons which decay to SM particles.  Multi-step cascades can occur naturally in hidden valley models \cite{Strassler:2006im,Han:2007ae}. In such models, production of the DM at terrestrial colliders and scattering in direct detection experiments can be generically suppressed by the small coupling between the dark and visible sectors. In contrast, indirect detection signals depend primarily on the annihilation rate of the DM to particles \emph{within} the dark sector; the small coupling between the sectors only suppresses the decay rate of the mediators to SM particles, which does not affect indirect searches provided the decay rate is small on astrophysical timescales (as in e.g. \cite{Rothstein:2009pm}). Thus cascade annihilations scenarios are often invoked to explain anomalies and suggest new DM signals.
For instance in \cite{Elor:2015tva,Hooper:2012cw,Ko:2014gha, Martin:2014sxa,Abdullah:2014lla, Berlin:2014pya,Cline:2014dwa,Liu:2014cma,Cline:2015qha} multi-step cascades were used to explain the apparent excess GeV gamma-rays identified in the central Milky Way \cite{Goodenough:2009gk,Hooper:2010mq,Boyarsky:2010dr,Hooper:2011ti,Abazajian:2012pn,Hooper:2013rwa,Gordon:2013vta,Huang:2013pda,Abazajian:2014fta,2014arXiv1402.6703D,Calore:2014xka, TheFermi-LAT:2015kwa}, while evading bounds from DM direct detection experiments.\footnote{There is recent evidence this excess may originate from a population of point-like objects, rather than DM \cite{Bartels:2015aea,Lee:2015fea}.} In general the injection of photons and other high energy secondary particles produced is constrained by a number of indirect searches. In particular we focus on:
\begin{itemize}
\item{Measurements of the CMB by {\it Planck} \cite{Ade:2015xua}}
\item{Bounds set by {\it Fermi} from DM searches in the Dwarf Spheroidal Galaxies of the Milky Way \cite{Ackermann:2015zua}}
\item{Measurements of the $e^+$ flux by AMS-02 \cite{Aguilar:2014mma,Accardo:2014lma}}
\end{itemize}

Constraints from the above three experiments can be parametrized model-independently for the case of direct DM annihilations (see for instance \cite{Cirelli:2008pk}), by classifying 
annihilations to all possible two-body SM final states, $\textrm{DM} + \textrm{DM} \rightarrow \textrm{SM} +\textrm{SM}$. For a given DM mass and final state, the spectra of secondary particles, is fixed independently of the form of the DM interaction and spin. Therefore constraints on DM annihilation rates are usually quoted in terms of the parameters relevant to the direct annihilation scenario, and do not encompass DM models embedded in a hidden sector.\footnote{\footnotesize{Signals and constraints for a class of 1-step hidden sector models were studies in~\cite{Mardon:2009gw}.}} Given the broad space of Hidden Sector DM models, it is essential to provide model-independent methods that cover the majority of model space. 

In the present work, we present DM mass dependent bounds on the DM cross section from the above three indirect detection experiments for DM annihilations via 0-6 step cascades to eight SM final states: $\gamma\gamma$, $e^+ e^-$, $\mu^+ \mu^-$, $\tau^+\tau^-$, $b \bar{b}$, $gg$, $W^+W^-$, and $h\bar{h}$. We remain agnostic about the details of the hidden sector, thus making our statements robust and model-independent.  Limits from the {\it Fermi} dwarfs and AMS-02 generally provide the strongest robust constraints on channels that are rich in photons and those that are not, respectively (although at sufficiently high masses, limits from H.E.S.S. \cite{Abramowski:2014tra} and VERITAS \cite{Zitzer:2015eqa} overtake those from \emph{Fermi}). While there may be arguably stronger bounds from the Galactic Center (e.g. \cite{Hooper:2012sr}) or galaxy clusters (e.g. \cite{2013ApJ...768..106S}), these limits depend strongly on the assumed DM density profile and/or the degree of substructure. We include the CMB limits because they are robust and almost independent of the spectrum of the annihilation products; thus we expect them to be nearly unaffected by the transition from 2-body to multi-body SM final states.
\newpage

Our results are presented in Fig.~\ref{fig:CMB0p3} - Fig.~\ref{fig:MasterLimits}, and our findings can be summarized as follows:
\begin{itemize}
\item{The {\it Planck} CMB bounds are robust and nearly model-independent varying by at most a factor of 1.5 over cascades with up to 6 steps for all final states.}
\item{For photon-rich final states (all states considered except electrons and muons), we find the dwarf limits yield the most sensitive robust constraint, and can be weakened or strengthened by about an order of magnitude or more as compared to the direct annihilation case. For high (low) DM masses and small (large) step number the dwarf bounds can be overtaken by the robust CMB bounds as the most limiting constraints.}
\item{For final states with few photons (electrons and muons), constraints from AMS-02 generally dominate the limits for low number of cascade steps. The limits can change by several orders of magnitude as compared to the direct case. As these weaken for higher DM masses and larger number of steps, CMB constraints become more important.}
\item{Taking the above three points into account we find that for a fixed DM mass and final state, the presence of a hidden sector can change the overall cross section constraints by up to an order of magnitude in either direction (although the effect can be much smaller).}
\end{itemize}
In addition to these constraints we also discuss how the bounds from multi-step cascades can be generalized to include the case of decays to $n$-body states in the dark sector. Finally as a supplement to this work we release code to generate the cascade spectrum. 

In Sec.~\ref{sec:Review} we review the procedure used in \cite{Elor:2015tva} to calculate the photon, electron and positron spectra from a multi-step cascade. Section~\ref{sec:Spec} contains a description of the SM final state spectra used. Then in Sec.~\ref{sec:nBody} we describe how results for multi-body decays can be estimated from our cascade results. Our main results are presented in Sec.~\ref{sec:CMB}-\ref{sec:AMS} where we show the model-independent bounds extracted from the CMB, dwarfs, and AMS-02 respectively. We briefly discuss estimated constraints from antiproton data in Sec.~\ref{sec:antiproton}. We discuss the interplay of the various experimental limits in Sec.~\ref{sec:gendis}, present an example application of these results to the Galactic Center excess in Sec.~\ref{sec:GCE}, and conclude in Sec.~\ref{sec:conclusion}. In the Appendices we describe the contents of the publicly available code, as well as additional details and cross-checks.

\begin{figure*}[t!]
\centering
\begin{tabular}{c}
\includegraphics[scale=0.47]{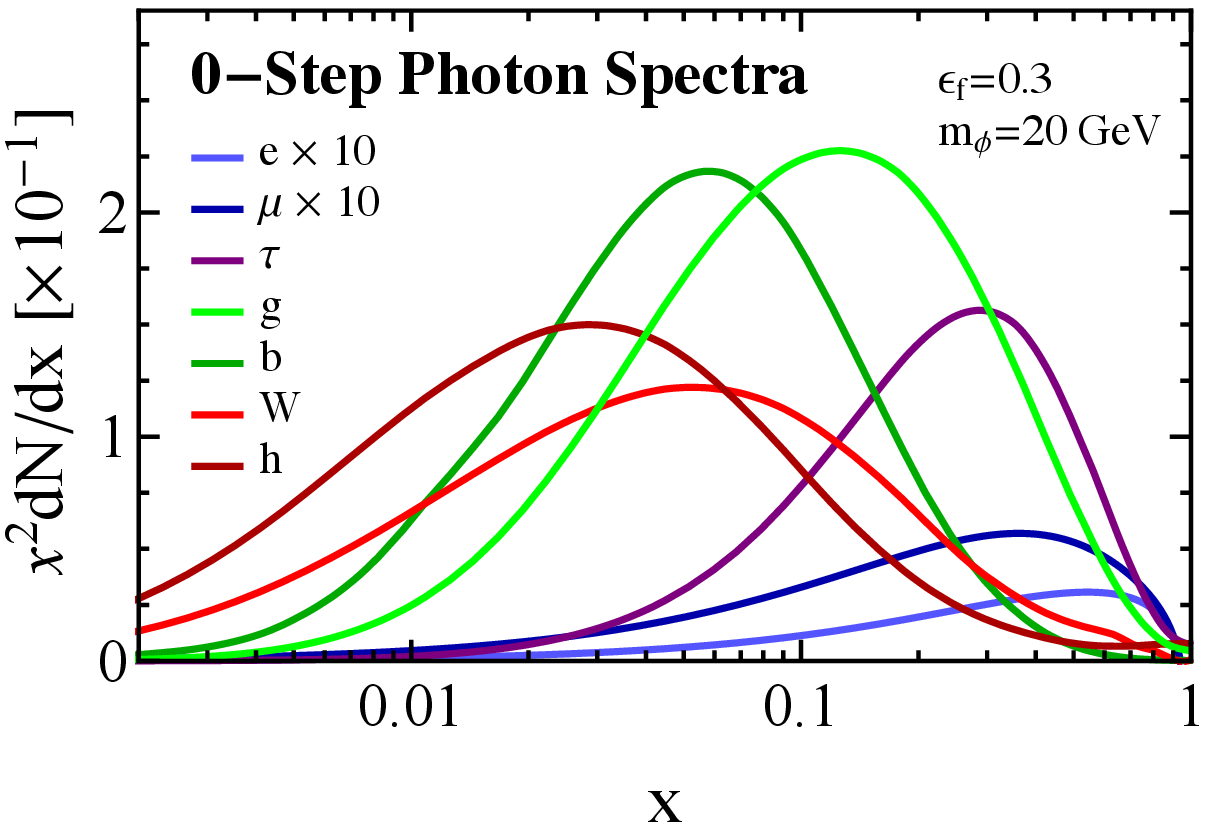} \hspace{0.05in}
\includegraphics[scale=0.47]{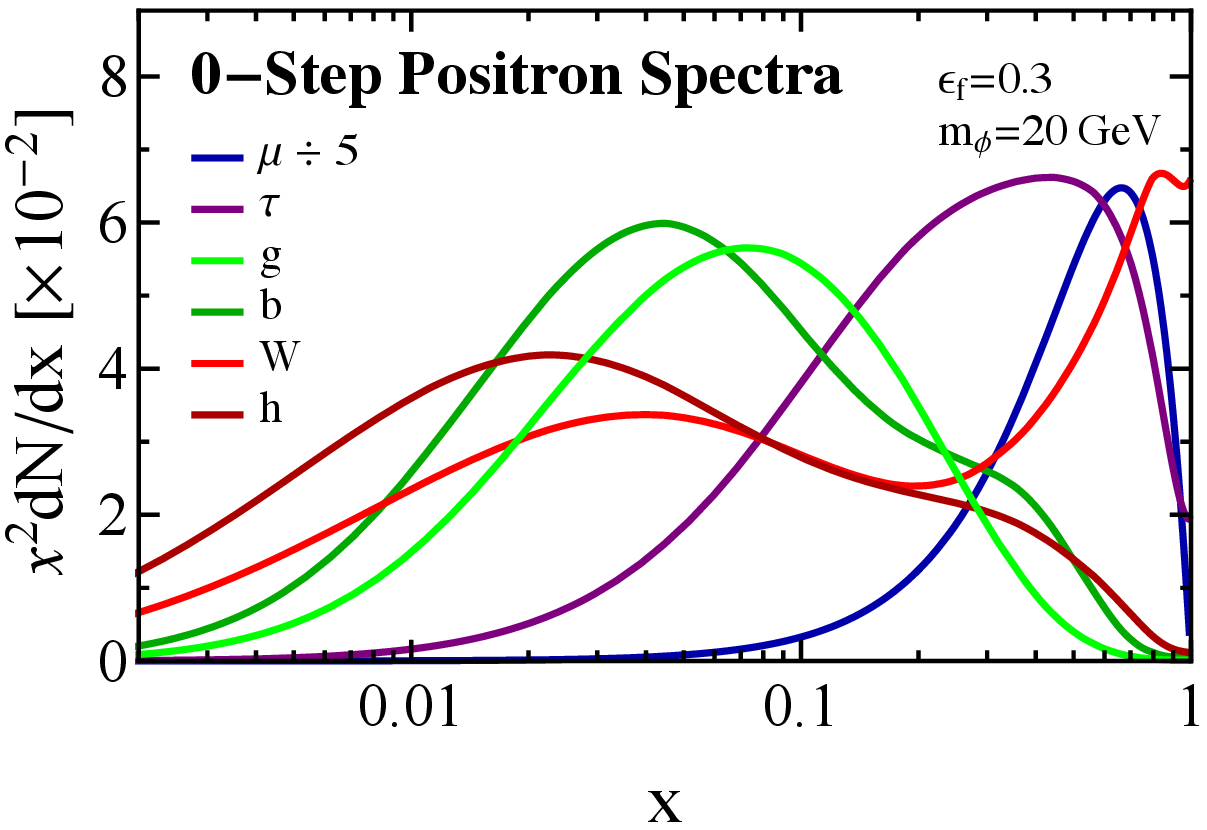} \hspace{0.05in}
\includegraphics[scale=0.47]{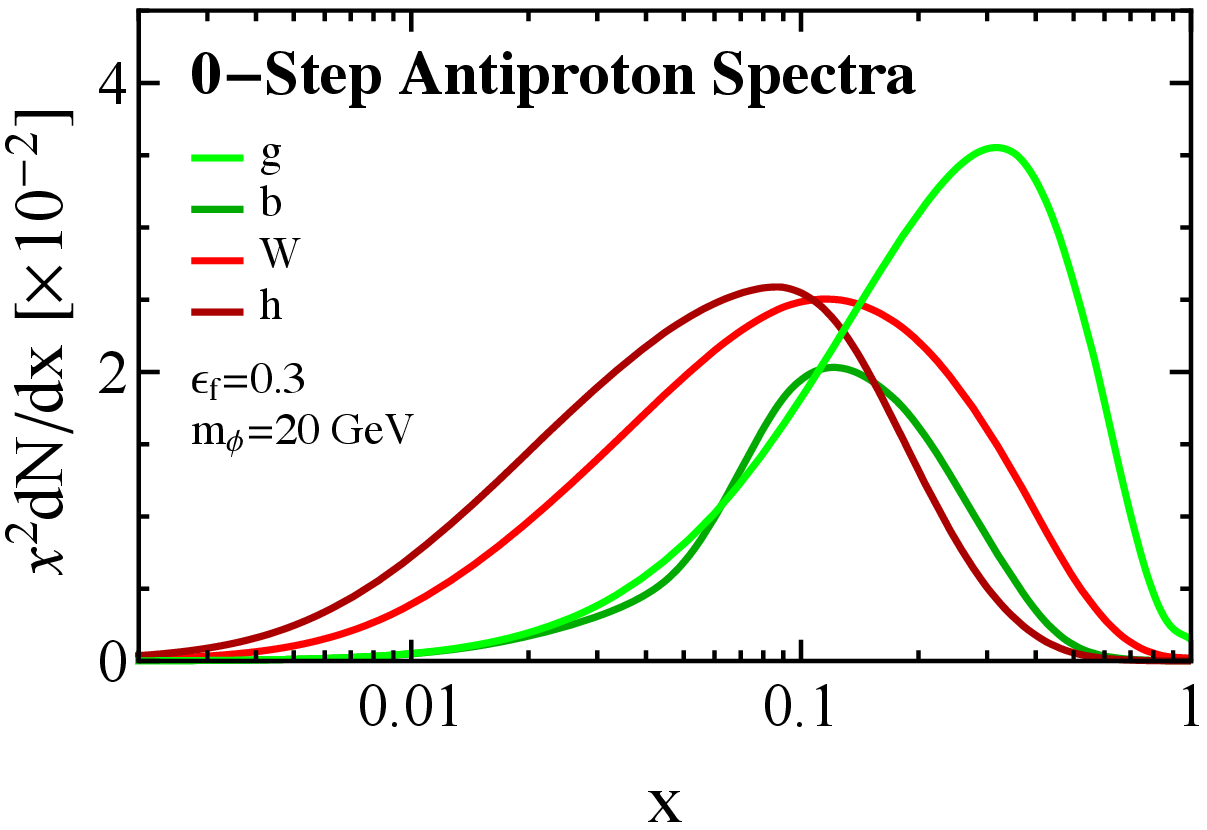}
\end{tabular}
\caption{\footnotesize{The 0-step or direct photon (left), positron (center) or antiproton (right) spectrum for the various final states considered in this work. We have left out the $\gamma \gamma$ spectrum in the photon case and the electron spectrum in the positron case as both of these are $\delta$-functions. Where applicable spectra are plotted with $\epsilon_f=0.3$ or $m_{\phi}=20$ GeV in the case of gluons.
}
}
\label{fig:0step}
\end{figure*}

\section{Multi-Step Cascade Annihilations}
\label{sec:Review}

The multi-step cascade annihilation scenario is illustrated schematically in the left panel of Fig.~\ref{fig:Cartoon}. In this setup the DM pair annihilates into two scalar mediators (the case of non-scalar mediators was discussed in \cite{Elor:2015tva} where the conclusions proved to be relatively insensitive to choice of vector or scalar mediator\footnote{\footnotesize{A thorough investigation of possible exceptions to this result is left to future work.}}) which subsequently decay through a (possibly) multi-step cascade in the dark sector, eventually producing a dark-sector state (with high multiplicity) that decays to the SM. Schematically we have:
\be\begin{aligned}
\chi \chi \rightarrow \phi_n \phi_n &\rightarrow 2 \times \phi_{n-1} \phi_{n-1} \rightarrow . . . \\
 &\rightarrow 2^{n-1} \times \phi_1 \phi_1 \rightarrow 2^{n} \times \text{(SM final state)}\,.
\label{eq:cascade}
\end{aligned}\ee
Here $n$ is the number of steps as defined above.

A variation on this picture occurs when any of the heavy hidden sector mediators goes off-shell and can therefore be integrated out yielding an effective vertex, now with a multi-body decay in the hidden sector of the form $\phi_n \to m \phi_{n-1}$, with $m>2$. This possibility is illustrated schematically in the right panel of Fig.~\ref{fig:Cartoon} and for a 1-step cascade the analogue to Eq.~\ref{eq:cascade} would be:
\begin{equation}
\chi \chi \to n \times \phi \to 2n \times \text{(SM final state)}\,,
\label{eq:nbodysetup}
\end{equation}
from which the extension to higher step cascades should be clear. Naively this framework seems like it could give quite different results to iterated 2-body decays. Yet in both cases, the main effect is to distribute the energy of the annihilation among a larger number of particles, thus increasing the multiplicity of the SM final state, lowering the average energy of the annihilation products, and broadening their spectrum. Consequently, limits on such scenarios can be broadly understood in terms of the $n$-step cascade results. This again highlights the point emphasized in \cite{Elor:2015tva} that the simple framework of $n$-step 2-body scalar cascades can describe a wide class of models and in this sense provide a relatively model-independent framework.

In Eq.~\ref{eq:cascade} and Eq.~\ref{eq:nbodysetup} ``(SM final state)'' denotes the SM particles produced by a single decay of $\phi_1$, which in turn will (in general) subsequently decay to produce observable photons, neutrinos and charged stable particles. For example a SM final state may produce additional photons due to final state radiation (FSR) or the decay of neutral pions produced during hadronization. The mass ratio between $\phi_1$ and the sum of the masses of the SM particles in this state, which we denote $\epsilon_f$ ($\epsilon_f \equiv ( \sum m_\mathrm{SM}) / m_1)$, controls the level of FSR and hadronization, and so is a useful parameters for describing these decays; the details are discussed in \cite{Elor:2015tva}. When the SM particles are massless, the relevant parameter is instead just the mass of the $\phi_1$, which we denote interchangeably as $m_1$ or $m_\phi$.

The spectrum of particles in an intermediate step of a cascade may be obtained using the method discussed in \cite{Elor:2015tva}, which we briefly review in this section. Consider the ``ith step'' decay $\phi_{i+1} \to \phi_i \phi_i$. In the rest frame of  $\phi_{i+1}$ we will denote the spectrum of the subsequent photons, electrons or positrons as $dN / dx_{i}$, where $x_i = 2E_i / m_{i+1}$, $m_{i+1}$ is the mass of $\phi_{i+1}$ and $E_i$ is the energy of the photon, electron or positron in the $\phi_{i+1}$ rest frame. We define $\epsilon_i = 2m_i/m_{i+1}$, and will (by default) assume a large mass hierarchy between cascades steps such that $\epsilon_i \ll 1$. Assume that the spectrum in the rest frame of the $\phi_{i}$ particle is known and denoted by $dN/dx_{i-1}$. In the limit of a large mass hierarchy the decay of $\phi_{i+1}$ produces two highly relativistic $\phi_i$ particles, each (in the rest frame of the $\phi_{i+1}$) carrying energy equal to $m_{i+1}/2 = m_i/\epsilon_i$. The photon, electron, or positron spectrum per annihilation in the rest frame of the $\phi_{i+1}$ is then given by a Lorentz boost, and takes the simple form
\be
\frac{dN}{dx_i} = 2 \int_{x_i}^{1} \frac{dx_{i-1}}{x_{i-1}} \frac{dN}{dx_{i-1}} + \mathcal{O}(\epsilon_i^2)\,.
\label{eq:boosteq}
\ee

In this way, we can begin with a direct spectrum of $dN/dx_0$ from $\phi_1 \to {\rm SM~final~state}$ -- the details of which are described in the next section -- and generate a cascade spectrum inductively. By repeated application of this formula we can see that the presence of each additional step in a cascade acts to broaden and soften the spectrum, and shift the peak to lower masses. Importantly the shapes of these cascade spectra are very simple, being characterized by just three pieces of information: the number of steps $n$, the SM final state (often denoted $f$), and the value of $\epsilon_f$. Such cascades are independent of the details of each of the intermediate steps, within the large-hierarchy ($\epsilon_i \ll 1$) approximation, and as such are independent of the various $\epsilon_i$.\footnote{The order of the error in the large-hierarchy approximation is $\epsilon_i^2$; see \cite{Elor:2015tva} for more details.}

As pointed out in \cite{Elor:2015tva}, although the large-hierarchy approximation seems to discard information, the more general case can be recovered quite easily. To see this, consider the opposite limit where $\epsilon_i \rightarrow 1$, so that $2 m_i \approx m_{i+1}$. In this case, the rest frames of the $\phi_{i+1}$ and $\phi_i$ are the same, so no boost needs to be applied. As such, in this ``degenerate limit'', the final spectrum of annihilation products is the same as that for a hierarchical cascade with one fewer step, with half the initial DM mass and half the annihilation cross-section. The intermediate regime, where neither $\epsilon_i$ nor $1-\epsilon_i$ are particularly small, smoothly interpolates between these two cases. Thus by studying the parameter space of $(m_\chi, \langle \sigma v \rangle$, no. of steps) in the hierarchical limit, it is possible to quickly estimate results for a general cascade. 

Again this framework is more general than it might initially appear. For example, simple extensions where a $\phi_i$ decays to two $\phi_{i-1}$ with different masses will not change our results in the large-hierarchy limit, as those results are independent of the intermediate masses. Additionally, as pointed out in \cite{Elor:2015tva}, for larger $n$ our cascade scenarios can approximate models with hadronization in the dark sector (see e.g. \cite{Freytsis:2014sua,vonHarling:2012sz}), and additionally as we will show in Sec.~\ref{sec:nBody}, multi-body decays can also be approximately captured within this framework.

Note that the cascade scenario must be physically self-consistent: the mass hierarchy between the DM mass and the SM particles in the final state must be sufficiently large to accommodate the specified number of steps. In other words, there is a hard upper limit on the number of steps allowed, for a given DM mass and final state. In detail, for an $n$-step cascade ending in a final state consisting of two particles each with mass $m_f$, we defined $\epsilon_f = 2 m_f/m_1$, $\epsilon_1 = 2m_1/m_2$, $\epsilon_2 = 2m_2/m_3$ and so on until $\epsilon_n=m_n/m_{\chi}$. Combining these, the DM mass is given in terms of $m_f$ and the $\epsilon$ factors by:
\be
m_\chi = 2^n \frac{m_f}{\epsilon_f  \epsilon_1  \epsilon_2  . . .  \epsilon_n}\,.
\label{eq:mDM}
\ee
If the $\epsilon_i$ factors are allowed to float, we can still say that $0 < \epsilon_i \leq 1$ in all cases (since each decaying particle must have enough mass to provide the rest masses of the decay products), setting a strict lower bound on the DM mass of:
\be m_\chi \geq 2^n m_f/\epsilon_f\,. \label{eq:kinematic} \ee 
Where this limit is \emph{not} satisfied, the spectra should not be thought of as potentially representing a physical dark-sector scenario, but only as a parameterization for general spectral broadening. For the massless final states considered here (photons and gluons) $m_f=0$, but we can still derive a condition from the value of $m_{\phi}$, specifically:
\be m_\chi \geq 2^{n-1} m_{\phi}\,. \label{eq:kinematicmphi} \ee 

\section{Direct Spectra}
\label{sec:Spec}

Using the formalism outlined in the previous section, from a given ``direct'' spectrum we can straightforwardly generate an $n$-step cascade spectrum, to compare with various indirect searches. We outline the different SM final states considered for the direct (0-step) spectra in this section. To obtain limits using bounds from the dwarfs, CMB and AMS-02 we need the spectrum of photons, electrons and positrons, and so we determine the spectrum for these particles arising from the boosted decays of the following eight SM states: $\gamma \gamma$, $e^+ e^-$, $\mu^+ \mu^-$, $\tau^+ \tau^-$, $\bar{b} b$, $W^+ W^-$, $h\bar{h}$, and $gg$.\footnote{In our publicly released code we also provide the antiproton spectrum for $b$-quarks, $W$-bosons, Higgs and gluons.} We choose these states as a representative sample of possible spectra. For example decays of light quarks generally give signals similar to those of $b$-quarks and the $ZZ$ final state is similar to $W^+ W^-$.

As discussed in the previous section, many of the cascade spectra depend on the parameter $\epsilon_f = \sum m_\mathrm{SM} / m_1 = 2m_f / m_1$ (the final equality holds for all the processes we consider here). In the context of generating the direct (0-step) spectrum, we can imagine two analogous processes: either the direct annihilation $\chi \chi \to {\rm SM~final~state}$, in which case $\epsilon_f = m_f/m_{\chi}$, or the final step in a cascade annihilation, $\phi_1 \to {\rm SM~final~state}$, so that $\epsilon_f = 2m_f/m_1$ as stated. If the $({\rm SM~final~state})$ is a photon or a gluon, then clearly $\epsilon_f$ is no longer a useful parameter; instead $m_{\phi} = m_1$ (equivalent to $2 m_\chi$ in the case of direct annihilation) plays this role. For many spectra no such parameter is needed. For example the $\gamma \gamma$ photon spectrum, as well as the positron spectra from $e^+ e^-$ or $\mu^+ \mu^-$ final states, are independent of any such parameter, since they are either just $\delta$-functions or arise from decay rather than FSR or hadronization. 

In all but five cases, we use the results of the PPPC4DMID package \cite{Cirelli:2010xx} to produce the direct spectra (hereafter referred to simply as PPPC). The exceptions to this are:
\begin{itemize}
\item the $\gamma \gamma$ photon and $e^+ e^-$ electron or positron spectra, which are just $\delta$-functions, to a good approximation (we neglect the effect of FSR on the $e^+ e^-$ spectra in the case of annihilation/decay to $e^+ e^-$),
\item the spectra of photons produced in conjunction with the $e^+ e^-$ and $\mu^+ \mu^-$ final states, for which we use the analytic results of \cite{Mardon:2009rc,Elor:2015tva} and \cite{Kuno:1999jp,Mardon:2009rc,Elor:2015tva} respectively,
\item the spectrum of electrons or positrons from muon decay, where we use the analytic Michel spectrum \cite{Michel:1949qe,Mardon:2009rc}.
\end{itemize}

Finally we briefly comment on the $\epsilon_f$ or $m_{\phi}$ dependence of the various direct spectra as it is often useful in interpreting results, noting that \cite{Elor:2015tva} has a more detailed discussion of several cases for photon spectra. For photons produced from $e^+ e^-$ and $\mu^+ \mu^-$ final states, the spectra arise entirely from FSR and so are strongly dependent on $\epsilon_f$, increasing in flux and becoming more sharply peaked near the maximum energy as $\epsilon_f$ decreases. Similarly the photon spectrum produced from the $W$-boson final state, in addition to a broad continuum peaked at low $x$, acquires a sharp spike at high $x$ due to FSR when $\epsilon_f$ becomes small. The photon spectrum from the $b$-quarks final state broadens and moves its peak to smaller $x$ as $\epsilon_f$ decreases; the gluon spectrum behaves similarly as $m_{\phi}$ increases. Finally the photon spectra from $\tau^+ \tau^-$ and $\bar{h}h$ final states are largely independent of $\epsilon_f$.

The positron spectra produced from the Higgs and tau final states again show no real variation with $\epsilon_f$. For positrons the spectrum from the $W^+ W^-$ final state is also quite independent of $\epsilon_f$, whilst the $b$-quark and gluon spectra behave much as they did in the photon case. Lastly, for antiprotons, once more the spectra from Higgs and $W$-boson final states are independent of $\epsilon_f$, whilst now for decreasing $\epsilon_f$ (increasing $m_{\phi}$) the $b$-quark (gluon) spectrum increases in height without substantially changing the position of its peak.

\newpage
\section{Multi-Body Cascades}
\label{sec:nBody}

\begin{figure}[t!]
\centering
  \includegraphics[scale=0.7]{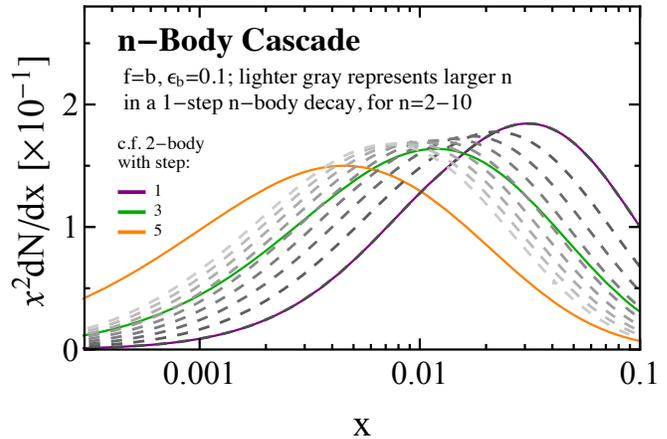}
  \caption{\footnotesize{The 1-step spectrum for an $n$-body cascade from a direct annihilation to $b\bar{b}$ with $\epsilon_b=0.1$ is shown as the dashed gray curves for $n=$ 2-10, where lighter curves correspond to larger $n$. In purple, green and orange we show a 2-body 1-step, 3-step and 5-step cascade spectrum respectively, for the same direct spectrum. These three curves outline the $n$-body results and show that the result of 1-step multi-body spectra should be encapsulated in the multi-step 2-body results.}}
  \label{fig:nbody}
\end{figure}

So far we have focused on cascades comprised of 2-body scalar decays. In this section we discuss the extension of this framework to the case of $n$-body cascades, schematically illustrated on the right of Fig.~\ref{fig:Cartoon}. As we will see, in the large hierarchies regime the $n$-body decays can be understood in terms of our existing 2-body results, again emphasizing the model-independence of our results. The explicit calculations and examples to help build intuition are provided in App.~\ref{app:multibody}. 

As explained in the introduction, a multi-body decay can arise if there is a heavy mediator in the cascade that has been integrated out. This can happen anywhere in a cascade, but here we restrict to a 1-step cascade of the form $\chi \chi \to n \times \phi \to 2n \times \text{(SM final state)}$ (c.f. Eq.~\ref{eq:cascade}). From here the extension to higher step cascades is intuitively clear, and in practice can be calculated using Eq.~\ref{eq:boosteq}. As shown in the appendix, an analogue of this equation can be derived for the multi-body case:
\begin{equation}
\frac{dN}{dx_1} = n(n-1)(n-2) \int_0^1 d \xi (1-\xi)^{n-3} \int_{x_1/\xi}^1 \frac{dx_0}{x_0} \frac{dN}{dx_0} + \mathcal{O}(\epsilon_1^2)
\label{eq:nbodyboosteq}
\end{equation}
where again $dN/dx_0$ represents the direct spectrum. Intuitively, the $dx_0$ integral accounts for the boosting of the decay products, just as in Eq.~\ref{eq:boosteq}, whilst the $\xi$ integral samples from the $n$-body phase space to give the correct degree of boosting.

At first glance it appears that this formula could produce marked differences to our standard cascade framework, but as we show in Fig.~\ref{fig:nbody}, this is not the case. There we show the 1-step spectrum for an $n$-body cascade ending in annihilation into the SM state $b \bar{b}$ with $\epsilon_b=0.1$, for $n=$ 2-10. Overlaid is the $m$-step 2-body cascade for $m=$ 1,3,5. Of course when $n=$ 2 and $m=$ 1 these two are the same by definition. But increasing $n$ and increasing $m$ perturb the spectra in quite similar ways (albeit to different degrees, as expected since the multiplicities of final-state particles are not equal for $m=n$ with $n > 2$), and so we expect the observational signatures of multi-body decays to lie within the space mapped out by the simple cascade annihilations. An example of the constraints on multi-body decays, and how they lie within the band of cascade constraints, is given in Fig.~\ref{fig:DwarfdeltaNBody}, in App.~\ref{app:multibody}.

\section{CMB Bounds from $\textit{Planck}$}
\label{sec:CMB}

DM annihilation during the cosmic dark ages can inject ionizing particles into the universe, modify the ionization history of the hydrogen and helium gas, and consequently perturb  the anisotropies of the CMB. Sensitive measurements of the CMB by $\textit{Planck}$ \cite{Ade:2015xua} (and previously WMAP and other experiments) can place quite model-independent limits on such energy injections, which when applied to DM annihilation are competitive with other indirect searches.

The figure of merit for CMB limits on DM annihilation is the parameter $p_\mathrm{ann} = f_\mathrm{eff} \langle \sigma v \rangle/m_\chi$, where $f_\textrm{eff}$ is an efficiency factor that depends on the spectrum of injected electrons and photons, and the other factors describe the total power injected by DM annihilation per unit time. In principle, different DM models could give rise to different patterns of anisotropies in the CMB -- but for WIMP models of DM that annihilate through $s$-wave processes, it has been shown \cite{Finkbeiner:2011dx} that the impact on the CMB is identical at the sub-percent level up to an overall rescaling by $p_\mathrm{ann}$ (related studies \cite{2011PhRvD..84b7302G,Hutsi:2011vx,2012JCAP...12..008G} independently found that the signal was largely controlled by a single parameter). In \cite{Slatyer:2015jla}-\cite{Slatyer:2015kla} this result was generalized to include any class of DM models for which $\langle \sigma v \rangle$ can be treated as a constant during the cosmic dark ages, which is generically true for the models considered in the present work.  We use the results of \cite{Slatyer:2015jla} to compute $f_{\text{eff}}$ as a function of DM mass and annihilation channel. In particular, we compute positron and photon spectra for the case of direct annihilations using PPPC, and then convolve to find the resulting spectrum for an $n$-step cascade as discussed above. The spectrum of electrons is equal to that of positrons by the assumption of charge symmetry. We then integrate the resulting spectra over the $f_{\textrm{eff}}(E)$ curves provided in \cite{Slatyer:2015jla} to obtain the weighted $f_{\textrm{eff}}(m_\chi)$ for $n=0$-6 step cascades for DM annihilations to various final states:
\bea \nonumber
f_{\textrm{eff}}(m_\chi) = \frac{\int^{m_\chi}_0 E dE \left[  2 f_{\textrm{eff}}^{e^+ e^-} \left(\frac{dN}{dE}\right)_{e^+} + f_{\textrm{eff}}^{\gamma} \left(\frac{dN}{dE}\right)_{\gamma}  \right]}{2m_\chi}\,. \\
\eea
We neglect the contribution from protons and antiprotons, as for all the channels we consider, the fraction of power proceeding into these channels is rather small, and consequently including them should only slightly increase $f_\mathrm{eff}$ \cite{Weniger:2013hja}. The constraints we present are therefore somewhat conservative (they could be strengthened slightly by a careful treatment of protons and antiprotons). As discussed in \cite{Slatyer:2015kla}, we use the best-estimate curves suited for the ``3 keV" baseline prescription, which are most appropriate for applying constraints derived by {\it{Planck}}.

The bound set on the annihilation parameter, $p_\mathrm{ann}$, from \textit{Planck} temperature and polarization data is taken to be \cite{Ade:2015xua}:
\bea
\frac{f_{\textrm{eff}} \langle \sigma v \rangle}{m_\chi} < 4.1 \times 10^{-28}~\textrm{cm}^3/\textrm{s}/\textrm{GeV}\,.
\label{eq:CMBbound}
\eea

In Fig.~\ref{fig:CMB0p3} we present our results for the bound on DM cross-section as a function of $m_\chi$ for various numbers of cascade steps and SM final states. We note that the number of steps does not affect the total power deposited by DM annihilation per unit time (at least in the simple scenario where all that power eventually goes into SM particles). Each additional step reduces the average energy of the final-state photons/positrons/electrons by a factor of 2, but simultaneously increases their multiplicity by a factor of 2. Thus the only possible impact on the constraints comes from the energy dependence of $f_\mathrm{eff}$, combined with the softening and broadening of the spectrum.

In accordance with our expectations, we find that the effect of the spectral broadening and softening is rather mild, typically changing the constraints by no more then 0.1-0.15 decades (corresponding to a factor of $\sim 1.5$). There is no general trend, in that constraints on these high-multiplicity final states may be either weaker or stronger than those pertaining to direct annihilation; this arises from the fact that $f_\mathrm{eff}$ is not a monotonic function of energy, so lowering the average energy of the injected particles may either increase or decrease the deposition efficiency. In general, $f_\mathrm{eff}$ and hence the upper bound on the ratio $\langle \sigma v \rangle/m_\chi$ varies less as a function of mass for higher-multiplicity final states (as expected, from the broader resulting spectrum), but this effect is very small. The choice of $\epsilon$ parameters, again, does not perturb the constraints outside this $\sim 0.15$-decade band. We refer the reader to the App.~\ref{app:CMBdetails} for additional details regarding the behavior of $f_{\textrm{eff}}$.

\section{Dwarf Limits from \textit{Fermi}}
\label{sec:Dwarfs}

The dwarf spheroidal galaxies of the Milky Way are expected to produce some of the brightest signals of DM annihilation on the sky. Whilst less intense than the emission expected from the galactic center, the dwarfs have the distinct advantage of an enormous reduction in the expected astrophysical background. These features make them ideal candidates for analysis with the data available from the \textit{Fermi} Gamma-Ray Space Telescope. Indeed the \textit{Fermi} Collaboration has set stringent limits on the DM annihilation cross-section using the dwarfs \cite{Ackermann:2015zua}, and together with the DES Collaboration have used 8 newly discovered dwarf satellites \cite{Bechtol:2015cbp,Koposov:2015cua} to set independent limits \cite{Drlica-Wagner:2015xua}. We note in passing that several groups have pointed out an apparent gamma-ray excess in the direction of one of the new dwarfs, Reticulum II \cite{Geringer-Sameth:2015lua,Drlica-Wagner:2015xua, Hooper:2015ula}, albeit with considerable variation as to its significance (with estimates ranging from $\sim 3\sigma$ to completely insignificant). We will not discuss this tentative excess here, other than to note as it appears roughly consistent with the emission coming from the GCE, the implications for dark sector cascades will be analogous to those discussed in \cite{Elor:2015tva}.

Here we focus on understanding how the presence of cascade annihilations can modify the limits obtained from these dwarf galaxies. In order to do this we use the publicly released bin-by-bin likelihoods provided for each of the dwarfs considered in \cite{Ackermann:2015zua}.\footnote{These results are available for download from \\ http://www-glast.stanford.edu/pub\_data/1048/} This analysis made use of 6 years of Pass 8 data and found no evidence for an excess over the expected background. Note the \textit{Fermi} collaboration produced an earlier analysis of the same dwarfs using 4 years of Pass 7 data in \cite{Ackermann:2013yva}. In App.~\ref{app:p7p8} we show that the results are similar between the two, but that the limits set using the newer analysis are usually about half an order of magnitude stronger.

Although \cite{Ackermann:2015zua} considered 25 dwarf galaxies, when setting limits they restricted this to 15, choosing a non-overlapping subset of dwarfs with kinematically determined $J$-factors. Specifically the 15 dwarfs considered were: Bootes I, Canes Venatici II, Carina, Coma Berenices, Draco, Fornax, Hercules, Leo II, Leo IV, Sculptor, Segue 1, Sextans, Ursa Major II, Ursa Minor, and Willman 1.

For a given dwarf \textit{Fermi} provides the likelihood curves as a function of the integrated energy flux in each of the energy bins considered in their analysis, covering the energy range from 500 MeV to 500 GeV. Thus to obtain the likelihood curves for our cascade models we need to firstly determine the integrated energy flux per bin. This will be a function of the DM mass $m_{\chi}$, annihilation cross-section $\langle \sigma v \rangle$, and shape of the cascade spectrum $dN/dx$ -- which itself depends on the number of cascade steps, the identity of the final state particle and possibly either $\epsilon_f$ or $m_{\phi}$. For an energy bin running from $E_{\rm min}$ to $E_{\rm max}$, the energy flux in GeV/cm$^2$/s is:
\begin{equation}
\Phi_E = \frac{\langle \sigma v \rangle}{8\pi m_{\chi}^2} \left[ \int_{E_{\rm min}}^{E_{\rm max}} E \frac{dN}{dE} dE \right] J_i\,,
\label{eq:Eflux}
\end{equation}
where $J_i$ is the $J$-factor appropriate for the individual dwarf $i$. Treating the energy bins as independent, we can simply multiply the likelihoods for the various bins to obtain the full likelihood for a given dwarf $i$: $\mathcal{L}_i\left({\boldsymbol \mu} \vert \mathcal{D}_i\right)$, which is a function of both the model parameters ${\boldsymbol \mu}$ and the data $\mathcal{D}_i$.  At a given mass and for a given channel (final state and number of cascade steps),  ${\boldsymbol \mu}$ just describes the annihilation cross-section $\langle \sigma v\rangle$. There is, however, one additional source of error that should be accounted for: the uncertainty in the $J$-factor. Following \cite{Ackermann:2015zua} we incorporate this as a nuisance parameter on the global likelihood, modifying the likelihood as follows:
\begin{equation}\begin{aligned}
&\tilde{\mathcal{L}}_i\left({\boldsymbol \mu}, J_i \vert \mathcal{D}_i\right) = \mathcal{L}_i\left({\boldsymbol \mu} \vert \mathcal{D}_i\right) \\&\times \frac{1}{\ln(10) J_i \sqrt{2\pi} \sigma_i} e^{-\left( \log_{10}(J_i) - \overline{\log_{10}(J_i)} \right)^2/2 \sigma_i^2} \,,
\label{eq:Jlike}
\end{aligned}\end{equation}
where for $\overline{\log_{10}(J_i)}$ and $\sigma_i$ we use the values provided in \cite{Ackermann:2015zua} for a Navarro-Frenk-White profile \cite{Navarro:1996gj}. This approach allows us to account for the $J$-factor uncertainties using the profile likelihood method \cite{Rolke:2004mj}. We obtain the full likelihood function by multiplying the likelihoods for each of the 15 dwarfs together.

Using this likelihood function, for a given DM mass and cascade spectrum we can then determine the 95\% confidence bound on the annihilation cross-section. We follow this procedure for cascade annihilations with 0-6 steps, for final state electrons, muons, taus, $b$-quarks, $W$-bosons, Higgses, photons and gluons, considering two different values of $\epsilon_f$ or $m_{\phi}$ where appropriate. 

Results are shown in Fig.~\ref{fig:DwarfLimits}. For the final states considered in \cite{Ackermann:2015zua}, our direct/0-step results are in agreement. Recall that there is a physical limitation on realizing a given cascade scenario set by $m_{\chi} \geq 2^n m_f/\epsilon_f$, as mentioned in Sec.~\ref{sec:Review}. The constraints corresponding to scenarios that satisfy this condition are indicated by darker lines, but we also show the limits for cases that do not satisfy this condition (and so cannot be physically realized as a cascade annihilation of the type we have considered), to demonstrate the effect of spectral broadening.

Before discussing results for each final state independently, there are a few generic features worth pointing out. Recall that higher-step cascades have a spectrum peaked at lower $x=E_{\gamma}/m_{\chi}$. Thus in order to produce emission at an equivalent energy, higher-step cascades require a larger DM mass, which in turn requires a larger cross-section to inject the same amount of power (as the DM number density scales inversely with the mass). Equivalently, at a fixed mass and cross-section, larger numbers of cascade steps will tend to produce a larger number of lower-energy photons; at low masses, some of these photons may lie outside the energy range of the \emph{Fermi} analysis, and the astrophysical backgrounds will also generally be larger at low energies. These factors tend to weaken the constraints, and indeed we see a systematic trend for weaker bounds with increasing $n$ for low-mass DM, for all channels.

Nevertheless this conclusion is not inevitable. Specific energy bins may allow stronger constraints than neighboring bins, purely due to statistical accidents; adding cascade steps smooths out such effects. The total number of emitted photons is increased with larger $n$ (albeit while preserving the total injected power). 

Most generically, if the DM mass is large, much of the spectrum may be above the 500 GeV cutoff of this analysis in the case of direct annihilation. In this case, adding cascade steps can strengthen the constraints by moving the photons into the range of sensitivity for the search. This effect is most pronounced, and occurs at the lowest DM masses, for final states with spectra peaked at large $x$ (electrons, muons, taus and photons): for softer direct-annihilation spectra, even at the heaviest masses tested, the peak of the spectrum does not move past 500 GeV. Inclusion of higher-energy data, e.g. from studies of the dwarf galaxies with VERITAS \cite{Zitzer:2015eqa}, would potentially strengthen the constraints at high DM masses, but for this reason we expect the improvement to be smaller for higher-step cascades.

Thus in general we see a weakening in the cascade constraints relative to the direct-annihilation case at low DM masses, and a strengthening at high DM masses, with the crossover point and the width of the band varying based on the SM final state. For some final states, the cascade constraints can be weaker or stronger than those for the direct-annihilation case by more than an order of magnitude. Let us now discuss the detailed results for each SM final state (shown in Fig.~\ref{fig:DwarfLimits}) separately:

\textit{Electrons:} the generic behaviors discussed above are clearly demonstrated in these results. There is also a striking difference between the results for direct and cascade annihilations. The photon spectrum in the direct case originates from FSR and is very sharply peaked (especially for small $\epsilon_f$); even a single cascade step will smooth out the spectrum and considerably change its shape. Further, the bounds are strongly dependent on the value of $\epsilon_f$, as the FSR photon spectrum diverges as $\epsilon_f \to 0$. As such, for smaller $\epsilon_f$ we expect stronger limits, and this is exactly what we observe. Nonetheless note that the position of the peak of the spectrum in $x$ is not strongly dependent on $\epsilon_f$, so we should expect the crossover behavior between different spectra mentioned above should happen at a similar location for different $\epsilon_f$ values and this is exactly what we observe. Finally note that the bumps in the direct spectrum are a result of the sharply peaked 0-step spectrum moving between energy bins. The width of these bumps is exactly the width of the energy bins in the data. As we move to cascade scenarios, the spectrum is smoothed out and the majority of the emission is no longer in a single bin, meaning these bumps vanish.

$\gamma\gamma$: the most noticeable feature here is the jagged direct spectrum. As the direct spectrum of $\gamma \gamma$ is just a $\delta$-function at the mass considered, these jumps are an extreme realization of the issue mentioned for the 0-step electron limits: we get a jump as the emission moves from one of the energy bins considered to the next. Of course physically the {\it Fermi} instrument has a finite energy resolution, which will act to smooth out such a sharp feature. To approximate this we smooth the 0-step by a Gaussian with a width set to 10\% of the energy value, yet this ultimately had little impact on the extracted limit. Note also that once the spike moves beyond 500 GeV, which occurs at roughly $\log_{10} m_{\chi} = 2.7$, the \textit{Fermi} data can no longer constrain this scenario so the limit completely drops off.

\textit{Muons:} the photon spectrum for the muon final state is very similar to that for the electron final state, except that it is slightly less dependent on $\epsilon_f$. The results here are accordingly very similar to those for the electron final state, except that the variations with $\epsilon_f$ are less pronounced.

\textit{Taus:} the fact that the tau spectrum is only weakly dependent on $\epsilon_f$ is clearly visible; otherwise only the generic behavior is apparent.

$b$-\textit{quarks:} There is a modest dependence on $\epsilon_f$, which does not change the qualitative results. The crossover  where the direct constraints become weaker than the cascade constraints occurs at a DM mass around 100 GeV. Due to the kinematic bounds, over the physically allowed region the variation in the band width is fairly modest varying by at most 0.4 decades.

\textit{Gluons:} the gluon spectrum behaves very similarly to the $b$-quark spectrum, if we swap decreasing $\epsilon_f$ for increasing $m_{\phi}$. As such the results are similar to those for $b$-quarks.

$W$-\textit{bosons:} firstly note that the kinematic edge in these results appears from the threshold requirement to have enough energy to create on-shell $W$'s. Other than this we see that the limits are somewhat stronger for smaller values of $\epsilon_f$, which is because the $W$ spectrum includes a small FSR component which is larger for smaller $\epsilon_f$. The width of the band of possible results is at most 0.7 decades. Again we also see a crossover where the direct constraints become weaker than the cascade constraints, here at roughly 500 GeV.

\textit{Higgs:} as with the $W$-bosons, our results again have a kinematic edge. Furthermore, like final state taus, the Higgs spectrum is only weakly dependent on $\epsilon_f$ and thus so are the results. As for the $W$ case, the width again has a maximum around 0.7 decades, whilst this time the direct crossover first happens at about a TeV. 

\section{Positron bounds from AMS-02}
\label{sec:AMS}

AMS-02 has recently released a precise measurement of cosmic ray electrons and positrons in the energy range of $\sim 1$ GeV to $\sim 500$ GeV~\cite{Aguilar:2014mma,Accardo:2014lma}.
The measured positron ratio exceeds the prediction of the standard cosmic ray 
propagation models at energies larger than $\sim 10$ GeV. There are many possible 
explanations for this rise in the positron ratio, including DM physics (although the annihilation scenario seems challenged by a range of other null results, e.g. \cite{Ade:2015xua}), nearby pulsars~\cite{Hooper:2008kg,Profumo:2008ms} 
or supernovae~\cite{Blasi:2009hv}.

The presence of an apparent large positron excess of unknown origin makes it challenging to set stringent limits on general DM annihilation scenarios. The situation is further complicated by the effects of solar modulation at energies 
below  $\sim 10$ GeV \cite{Gleeson:1968zza,Moskalenko:2001ya,Beischer:2009zz}, which modifies the cosmic ray flux in a charge-dependent manner and adds significant astrophysical uncertainties. However, the data do indicate that both the positron ratio (flux of $e^+$ divided by the flux of $e^+ + e^-$) and the fluxes of cosmic ray electrons and positrons are fairly smooth; there is no clear structure 
in the spectrum within the energy resolution of AMS-02. Accordingly, it is possible to set quite strong constraints on DM models that predict a sharp spectral feature in the positron spectrum (e.g. \cite{Bergstrom:2013jra,Hooper:2012gq,Ibarra:2013zia}).

As discussed in the previous sections, DM annihilation through multi-step cascades usually gives rise to a softer and broader
spectrum than direct annihilation to the SM states, generally leading to weaker
bounds from AMS-02. In this section, we study this effect quantitatively. We note that our goal here is to study the impact of these spectral changes, not to explore possible explanations for the rise in the positron fraction or systematic uncertainties in the modeling of the background or signal.

To set bounds on annihilating DM, we first need to parametrize or model 
the backgrounds. Here the backgrounds that we refer to are the astrophysical 
cosmic ray flux, plus some new smooth ingredient to account for the observed rise in the positron flux. Since we do not know the source of the new ingredient,  
polynomial functions of degree 6 are introduced to fit the AMS-02 positron flux data (the 6 degrees are employed to obtain a good $\chi^2$ fit to the data).
To derive the limits, we float the 6 parameters from the polynomial functions 
within 30$\%$ of the best fit values from the fit without DM, together with the DM annihilation cross-section. We check that increasing the range of allowed values for the background parameters does not weaken the constraints.

We derive limits from only the positron flux, as both the positron and electron backgrounds are required to float in the fit to the positron ratio. Such an analysis would require many additional free parameters, and is beyond the scope of the current work. As a cross-check we attempted a simplified fit to the positron ratio data (using AMS-02 measurements of the total $e^++e^-$ spectrum) and found constraints of comparable strength to those we present here.

The positron flux from DM annihilation is obtained by propagating the injected
positron spectrum using the public code DRAGON~\cite{Evoli:2008dv,2011ascl.soft06011M}. There are substantial
systematic uncertainties in the propagation of electrons and positrons in the 
galaxy, affecting diffusion, energy loss, convection, and solar modulation. 
In particular, accounting for uncertainties in the modeling of energy loss and solar modulation can significantly weaken the constraints on DM annihilation. 

\newpage

Once electrons and positrons are injected into the halo, they will diffuse and lose
energy. Their number density $N_i$ evolves according to the following diffusion equation,
\begin{eqnarray}
   \frac{ \partial N_i} { \partial t} &=&
      \vec{\nabla} \cdot \left( D \vec{\nabla} - \vec{v}_c \right) N_i
      + \frac{\partial} { \partial p} \left ( \dot{p}
      -\frac{p}{3} \vec{\nabla} \cdot \vec{v_c} \right) N_i
   \nonumber \\
   &+& \frac{ \partial} { \partial p} p^2 D_{pp} \frac{\partial}{\partial p}
      \frac{ N_i} { p^2}  +   Q_i ( p, r, z)
   \nonumber \\
     &+&  \sum_{j>i} \beta n_\mathrm{gas} (r,z) \sigma_{ji} N_j
      - \beta n_\mathrm{gas} \sigma_i^{in}( E_k) N_i ~,
   \label{eqn::prop}
\end{eqnarray}
where $D$ is the spatial diffusion coefficient, depending on the spatial position
and energy. It is parametrized by the following form
\begin{equation}
   D ( \rho, r, z) = D_0 \mathrm{e}^{|z|/z_t} \left( \frac{ \rho} 
      {\rho_0} \right)^\delta~,
\end{equation}
where we assume the diffusion zone is axisymmetric, and use the cylindrical coordinate system $( r, z) $. Most of the electrons and positrons are trapped in the 
diffusion zone with thickness $2 z_t$. Here $\rho = p / (Ze)   $ is the rigidity of the charged
particle with $Z= 1$ for electron and positron. 
$D_0$ normalizes the diffusion at the 
rigidity $\rho_0 = 4$ GV. In Eq.~\ref{eqn::prop}, $v_c$
is the velocity of the convection winds; $\dot{p}$ accounts for the energy loss; 
$Q_i$ is the source of the cosmic ray, where DM is one kind of the source; 
$D_{pp}$ accounts for  the diffusion in the momentum space; the last two terms
in Eq.~\ref{eqn::prop} parameterize how the nuclei inelastic scattering with the gas to 
affects the number density of the cosmic rays. Although there are many parameters
in the diffusion equation, we do not simulate the backgrounds (instead modeling them with a polynomial function), which decreases the systematic uncertainties of the limit substantially.

\begin{figure*}[t!]
\centering
\begin{tabular}{c}
\includegraphics[scale=0.7]{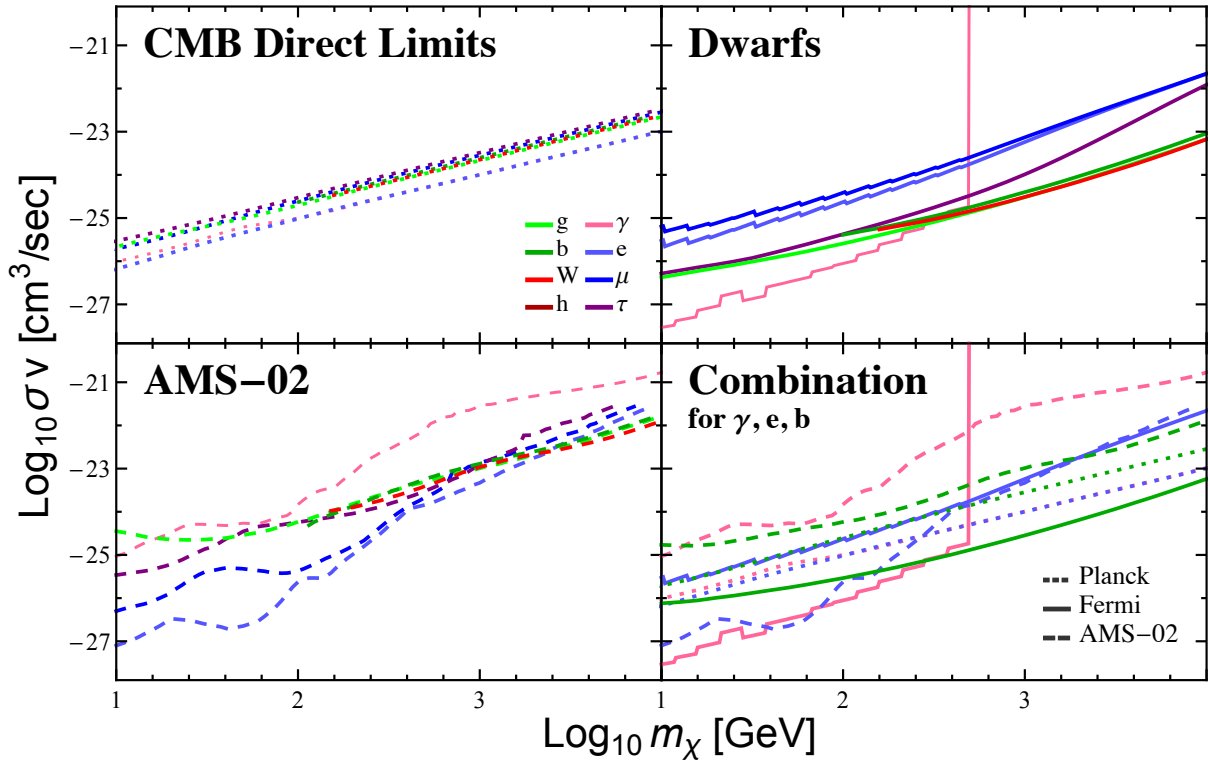}
\end{tabular}
\caption{\footnotesize{Constraints (at $95\%$ confidence) on $\langle \sigma v \rangle$ for the case of direct annihilations to photons, electrons, muons, taus, $b$-quarks, gluons, $W$'s and Higgs final states derived from CMB (top left), dwarfs (top right) and AMS-02 (bottom left). In the bottom right panel we overlay the constraints from all three experiments for the case of direct annihilations to final state photons, electrons and $b$-quarks.}}
\label{fig:DirectCombo}
\end{figure*}

We use a specific model to propagate the electrons and positrons 
injected by DM annihilation~\cite{Evoli:2011id}. In this model, $D_0 = 2.7 \times 10^{28}$ cm$^2$/s, 
$z_t = 4$ kpc, $\delta = 0.6$ and we take the local density of DM to be 0.4 GeV/cm$^3$ and the density profile
to be the Navarro-Frenk-White profile. We set the convection and diffusion in momentum 
space equal to zero, since they do not change the spectrum significantly in the 
energy range of interest~\cite{Strong:2007nh}. There are other propagation 
models with different diffusion terms or diffusion zone heights that can be employed 
here.
However, since the energy loss effect is dominant for the propagation of high 
energy leptons, we choose only one propagation model to derive the limits. While there may be remaining systematic effects due to the choice of propagation model, we reiterate that the purpose of this analysis is not to explore all the uncertainties in these constraints.

Cosmic-ray propagation is affected by the magnetic field, which determines both how the cosmic rays diffuse and their energy losses due to synchrotron radiation. The magnetic field is modeled by two components, one regular and one 
turbulent (irregular) ~\cite{Pshirkov:2011um, DiBernardo:2012zu}. 
\cite{Jaffe:2009hh} gives the constraints on these components. To be 
conservative, here we set the value of the magnetic field at the Sun to
$B_\odot \sim 8.9$ $\mu$G. With this magnetic field, the local radiation field
and magnetic field energy density is 3.1 eV/cm$^3$, which is 
close to (but somewhat higher than) the 2.6 eV/cm$^3$ value used for conservative constraints in \cite{Bergstrom:2013jra}. For this reason, 
the constraint we obtain for the direct annihilation is slightly weaker than even the conservative case studied in \cite{Bergstrom:2013jra}, as the energy loss rate for the positrons is higher. The main effect of changing the local energy density is to rescale all the constraint curves, with lesser effects on the variation of the constraint with DM mass and number of cascade steps. 

For cosmic rays with energy smaller than $10$ GeV, although there are 
many other parameters in the propagation model, 
we only consider the uncertainty from the solar modulation, which is modeled by
the modulation potential. The modulation potential $\phi$ in the range of 
$( 0, 1 )$ GeV is fixed by minimizing the $\chi^2$ to fit the AMS-02 data.  

\begin{figure*}[htbp]
\vspace*{+0.27cm}
\hspace*{-0.9cm}
\includegraphics[scale=0.8]{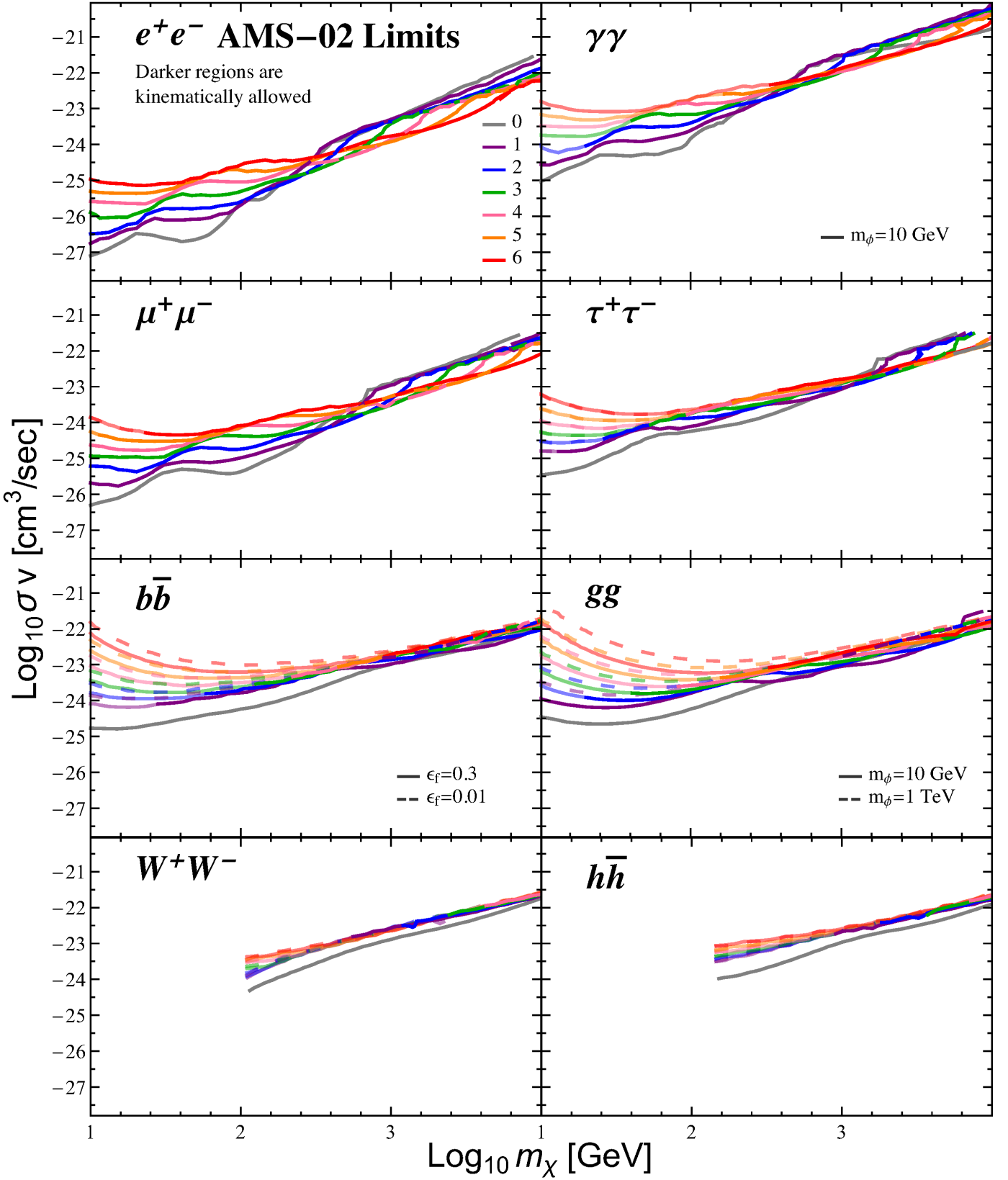}
\caption{\footnotesize{95\% confidence limits on DM cross-section for cascade
models. Details are similar to the previous two plots. The limits obtained are strongest for electron and muon final states, and generically we find that the addition of cascades steps can change the limits by up to several orders of magnitude.}}
\label{fig:AMSLimits}
\end{figure*}

In summary, we derive the limits on DM annihilation by using AMS-02 positron 
flux starting from $1 \mathrm{GeV}$. The background is parametrized by a polynomial function of 6 degree, and
to derive the bounds we let the 6 parameters float within 30$\%$ of their best 
fit values. The diffusion model is employed here to propagate the DM positron flux,
and the solar modulation potential is allowed to float in the range  
$( 0, 1 )~\mathrm{GeV}$ when minimizing the likelihood function. The limits are summarized in Fig.~\ref{fig:AMSLimits}.

In general, similar to the dwarf galaxies, the constraints on cascade models can be substantially weaker than those on the direct-annihilation case for low DM masses (below $\sim 100$ GeV), by up to several orders of magnitude depending on the channel. This weakening likely arises from a combination of (a) positrons falling below the minimum energy of the search, and (b) broadening of the spectrum making it easier for the background model to compensate for a DM component. The effect can be up to two orders of magnitude in most channels at sufficiently low masses (the exceptions are the $W$, Higgs and $b$-quark final states where low DM masses are kinematically forbidden). At high masses, the bands of possible constraints are narrower, of order half an order of magnitude or less; for the electron, muon, tau and gamma final states the direct-annihilation constraints are systematically weaker than those for cascade scenarios. This is likely due to the cascade scenarios producing greater numbers of positrons in the energy range of the search, but may also be due to the hardening of the positron flux at high energies mimicking a hard signal from DM annihilation.

\section{Antiproton ratio bounds}
\label{sec:antiproton}

Measurements of cosmic ray antiprotons may also provide stringent constraints on DM annihilation, especially to hadronic final states (e.g. \cite{Fornengo:2013xda, Cirelli:2013hv, Evoli:2015vaa}). However, as with the positron data, uncertainties in cosmic ray propagation and the modeling of the background cosmic ray population can substantially affect sensitivity. Here we present a brief discussion of the dependence of antiproton bounds on the number of cascade steps; we caution that this analysis does not include a comprehensive study of uncertainties in the background and propagation models, although we do study two alternative propagation models. Since it is difficult for antiproton limits to robustly improve on constraints from \emph{Fermi} observations of dwarf galaxies (unlike the positron bounds, which have sensitivity to channels that produce very few photons), we do not include these limits in our final combined constraint plots. In this section we employ data from BESS~\cite{Orito:1999re,Asaoka:2001fv}, CAPRICE~\cite{Boezio:2001ac}, and 
PAMELA~\cite{Adriani:2012paa}.\footnote{Preliminary antiproton data have been presented by AMS-02, but we do not employ these results; we reiterate that the purpose of this section is not to provide a robust new bound on DM annihilation from antiprotons (as this requires an in-depth study of systematic uncertainties that is beyond the scope of this work), but to investigate broadly how these bounds may be expected to vary for multi-step cascades.}

The propagation of antiprotons and protons follows the transport equation, as given in Eq.~\ref{eqn::prop}. The parameters in the propagation models are derived by reproducing the B/C data, following the approach in \cite{DiBernardo:2009ku,Evoli:2011id,Urbano:2014hda,Donato:2001ms,Moskalenko:2001ya}. 
In this section, we will use the KRA and THN models~\cite{Urbano:2014hda} as illustrative examples of possible propagation scenarios; the THN model yields more conservative constraints on DM annihilation, since it possesses a thin diffusion zone ($z_t = 0.5$) and small diffusion coefficient ($D_0 = 0.32 \times 10^{28}$ cm$^2$/s). We assume that the DM density is well described by the Navarro-Frenk-White profile, and that the local DM density is $0.4$ GeV/cm$^3$. 

Instead of parametrizing the astrophysical background by polynomial functions, the astrophysical background of proton and antiproton are simulated by DRAGON.
Besides the propagation parameters, we must also determine the solar modulation parameter $\phi$. Since the different experiments whose data we use operated at different times, each one must be allowed to have a different value of this parameter. We determine $\phi$ for each experiment by minimizing the $\chi^2$ considering only the ratio $\bar{p}/{p}$; we then re-fit the data including a DM component but holding $\phi$ fixed.

As in Sec.\ref{sec:AMS}, we employ DRAGON~\cite{Evoli:2008dv,2011ascl.soft06011M} to propagate the flux from DM annihilation. Following \cite{Urbano:2014hda}, we derive our bounds from the antiproton-to-proton ratio data, to reduce systematic uncertainties and allow combination of results from different experiments.

\begin{figure*}[t!]
\centering
\includegraphics[scale=0.7]{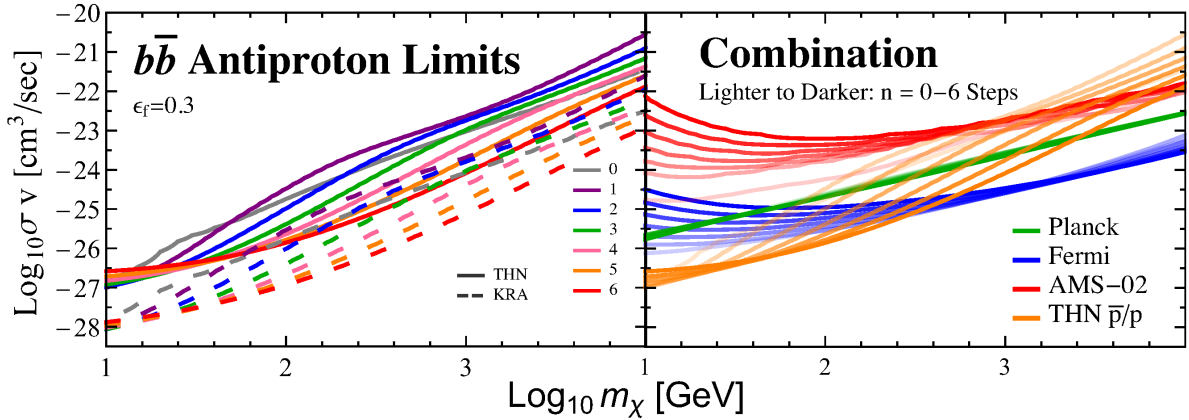}
\caption{The $95\%$ confidence limit bound on $\langle \sigma v \rangle$ for $n = 0-6$ step cascades from antiproton fraction measurements, for the $b \bar{b}$ final state with 
$\epsilon_f = 0.3$. The left panel shows the limits for the THN and KRA propagation models; the right panel reproduces the THN bounds, overlaid with the other limits discussed in this work.}
\label{fig:antip}
\end{figure*}

The constraints on the DM annihilation cross section, for final states comprised of $b$ quarks, are summarized in Fig.~\ref{fig:antip}. For this figure we set $\epsilon_f = 0.3$; we also include the case of direct annihilation. We see that the choice of propagation model changes the constraints by up to an order of magnitude, with the constraints from the THN model being weaker.

We find that increasing the cascade step number strengthens these antiproton bounds for DM masses larger than $\sim 20 \mathrm{GeV}$. The reason is that the $\bar{p}/{p}$ data are considerably more precise (and abundant) at energies smaller than $10~\mathrm{GeV}$; increasing the number of steps, for a given DM mass, shifts the antiproton spectrum down in energy, and into this better-constrained region.

It is also worth noting that in this case, the direct annihilation bound does not behave like a limiting case of the bounds on an $n$-step cascade. This is because by holding $\epsilon_f$ fixed for the $n$-step cascade, we are fixing the center-of-mass energy for the $b$ quark production (i.e. the decay of the last mediator in the chain); in contrast, for the direct annihilation this quantity is determined by the DM mass. The antiproton production is sensitive to the details of the hadronization process and requires a relatively large center-of-mass energy; consequently the antiproton spectrum is quite dependent on the center-of-mass energy for the final decay/annihilation, in contrast to the electron and photon spectra. This suggests that our model-independent cascade formalism is likely to be a worse approximation for antiproton final states, with generically stronger dependence on $\epsilon_f$, compared to the other channels we have considered.

Nonetheless, in the case we have studied, even the conservative THN model admits constraints on high-step cascades that are stronger than those from the \emph{Fermi} dwarf observations for relatively light DM (e.g. for DM masses below $\sim100$ GeV, for $n \ge 4$). This suggests that improved antiproton data and better characterization of the background and cosmic ray propagation could be particularly important in constraining cascade-type models with hadronic final states.

\section{General Discussion}
\label{sec:gendis}

\begin{figure*}[htbp]
\vspace*{+0.18cm}
\hspace*{-1.1cm}
\includegraphics[scale=0.8]{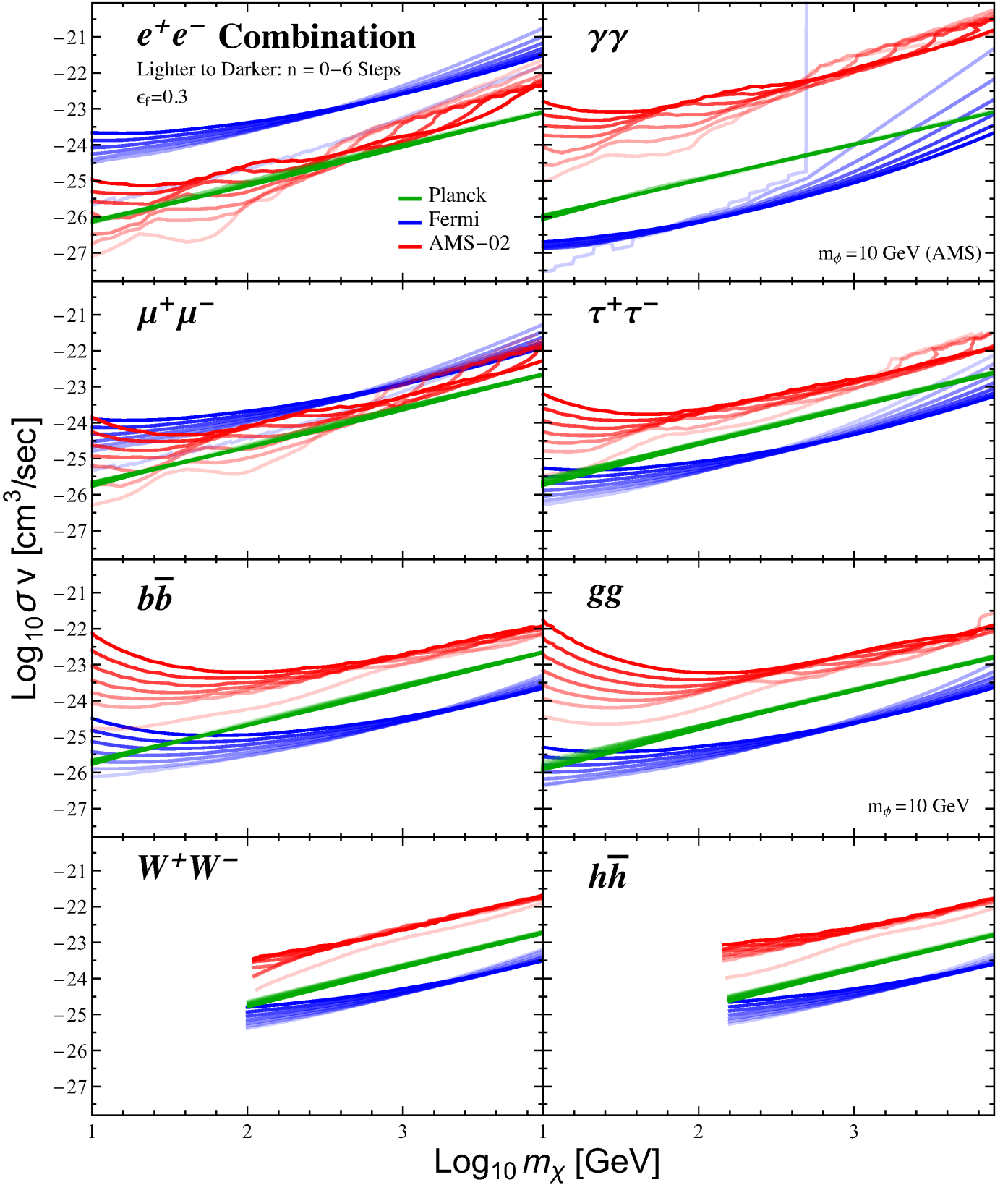}
\caption{\footnotesize{Overlaid constraints ($95\%$ confidence) from the CMB (green), AMS-02 (red) and the $\it{Fermi}$ Dwarfs (blue) for $n =$ 0-6 step cascades (lighter to darker shading) for various SM final states. The CMB bounds are very weakly dependent on the number of cascade steps, while the AMS-02 and dwarf results change noticeably. The AMS-02 bounds are most constraining for electron and muon final states, and weaken rapidly as the positron spectrum broadens with increasing cascades steps. The dwarfs are generally most constraining for final states with a high multiplicity of photons.}}
\label{fig:MasterLimits}
\end{figure*}

We summarize our main results in Fig.~\ref{fig:MasterLimits}, where we overlay the combined constraints from the three experiments as a function of DM mass for an $n =$ 0-6 step hidden sector cascade. Furthermore in Fig.~\ref{fig:DirectCombo} we show results just for the direct, or 0-step, annihilation, in order to highlight the interplay between the experiments. As discussed above the CMB constraint is fairly insensitive to the SM final state and number of cascade steps. The AMS-02 bounds, which are most constraining for direct annihilation for electron and muon final states, weaken rapidly at low masses as the positron spectrum broadens with increasing cascade steps, but for masses above a few hundred GeV are also fairly robust. The dwarfs are generally most constraining for final states with a high multiplicity of photons. However, for lepton-rich or photon-rich final states respectively, the CMB bounds can become more constraining than the AMS-02 or \emph{Fermi} limits for large numbers of cascade steps at low masses, or small numbers of cascade steps at high masses. We summarize the results the various SM final states below. 

\textit{Electrons:}  The spectrum of positron and photon spectrum is very sharply peaked in the case of direct DM annihilations to $e^+ e^-$. Thus AMS-02 places the most constraining bound for $n =$ 0-4 step cascades at low masses $m_{\chi} \lesssim 400 \mathrm{GeV}$. As the number of steps increases, $n > 3$, the spectrum smooths and broadens thereby weakening the AMS-02 bound so that the CMB bound becomes the most constraining. The CMB bounds are generically stronger at high DM masses, above a few hundred GeV. The dwarf limits are, in all cases, 1-2 orders of magnitude less constraining than the AMS-02 and CMB bounds. This is unsurprising given the dwarfs are only sensitive to the photon spectrum from the final state electrons, which represents only a small fraction of the available power per annihilation.

$\gamma\gamma$: The strongest constraints almost always arise from the \emph{Fermi} dwarfs, although at high DM masses and for small numbers of steps, the CMB bounds may be more stringent. However, in this case VERITAS or H.E.S.S dwarf searches may actually provide a stronger limit. For AMS-02 the positron spectrum is similar in shape to that of the electron channel; the photon generates a hard electron spectrum via Drell-Yan. Nonetheless this process is suppressed by a factor of $\alpha_e$ as well as phase space. Combining these, approximately two order magnitude suppression relative to electron case would be expected and is in fact observed.

\textit{Muons:} Recall that the spectrum of positrons and photons from DM annihilations to muons is similar to the corresponding spectra for the electron final state, except the photon spectrum is less dependent on $\epsilon_f$ and the positron spectrum is somewhat broader. For 0-2 cascade steps, the most stringent constraints are from AMS-02 at low masses, below a few hundred GeV. At higher step numbers (for all masses) and higher masses (for all cascade scenarios), the CMB limit becomes more restrictive.

\textit{Taus:} The tau final state is richer in photons than the other leptonic final states, and yields smoother and broader photon and positron spectra even in case of direct annihilation. Thus the bound from the dwarfs is more sensitive and constraining than AMS-02, and generally also stronger than the CMB limits. The exceptions are at low mass and large number of steps, or inversely high mass and a small number of steps, as in both cases the CMB bounds dominate the constraint.

$b$-\textit{quarks:} The direct spectrum for DM annihilations to $b\bar{b}$ is much softer then the previously discussed channels. So the \emph{Fermi} dwarf limits almost always provide the strongest constraint; for low masses and $n =$ 3-6 steps there is a region of parameter space where the CMB bounds appear to be more stringent, however this region is kinematically disallowed.

\textit{Gluons:} As previously discussed the gluon spectrum behaves very similarly to the $b$-quark spectrum, if we swap the decreasing $\epsilon_f$ for increasing $m_{\phi}$. As such the results are similar to those for $b\bar{b}$.

$W$-\textit{bosons:} For annihilation to $W$ final states the bounds are quite robust, with the dwarfs always setting the strongest limits. 

\textit{Higgs:} Annihilations to final state Higgses is similar to the $W$ case; the results are almost identical aside from the difference in the kinematic edge between the H and $W$ mass.

Finally let us say a few words about how these results are likely to change in the near future as additional experimental data becomes available. For the CMB limits, the shape of the limits is dictated by processes occurring in the early universe -- additional data from {\it Planck} will strengthen the limits but not change their qualitative behavior. The situation is similar for the dwarf limits. In the absence of a signal the shape of our results are being determined by three factors: the DM spectrum, the spectrum of the backgrounds, and the response of the {\it Fermi} detector. These will not change with additional data.\footnote{A caveat to this would be if {\it Fermi} again reprocesses their entire dataset and detector response, as they did in the move from Pass 7 to Pass 8. Nonetheless as we explore in App.~\ref{app:p7p8} even this does not greatly impact the qualitative results.} Again additional data and especially the addition of new dwarfs will just strengthen the limits (unless, of course, there is a detection). Given AMS-02 is still relatively early in its mission cycle, it is possible the relative sensitivity at different energies could change as more data is collected, which could alter the behavior of the constraints with respect to the number of cascade steps; however, the main effect is likely to be just to strengthen the limits. The addition of AMS-02 antiproton data to our estimated antiproton bounds could likewise impact our results given the different systematic effects (compared to earlier experiments) and improved sensitivity at higher energies; further AMS-02 observations might also admit improvements to our positron and antiproton constraints (and changes in the relative sensitivity to multi-step cascades) via improvements in the background and signal modeling.

\section{An Example Application: The Galactic Center Excess}
\label{sec:GCE}

\begin{figure*}[t!]
\centering
\begin{tabular}{c}
\includegraphics[scale=0.7]{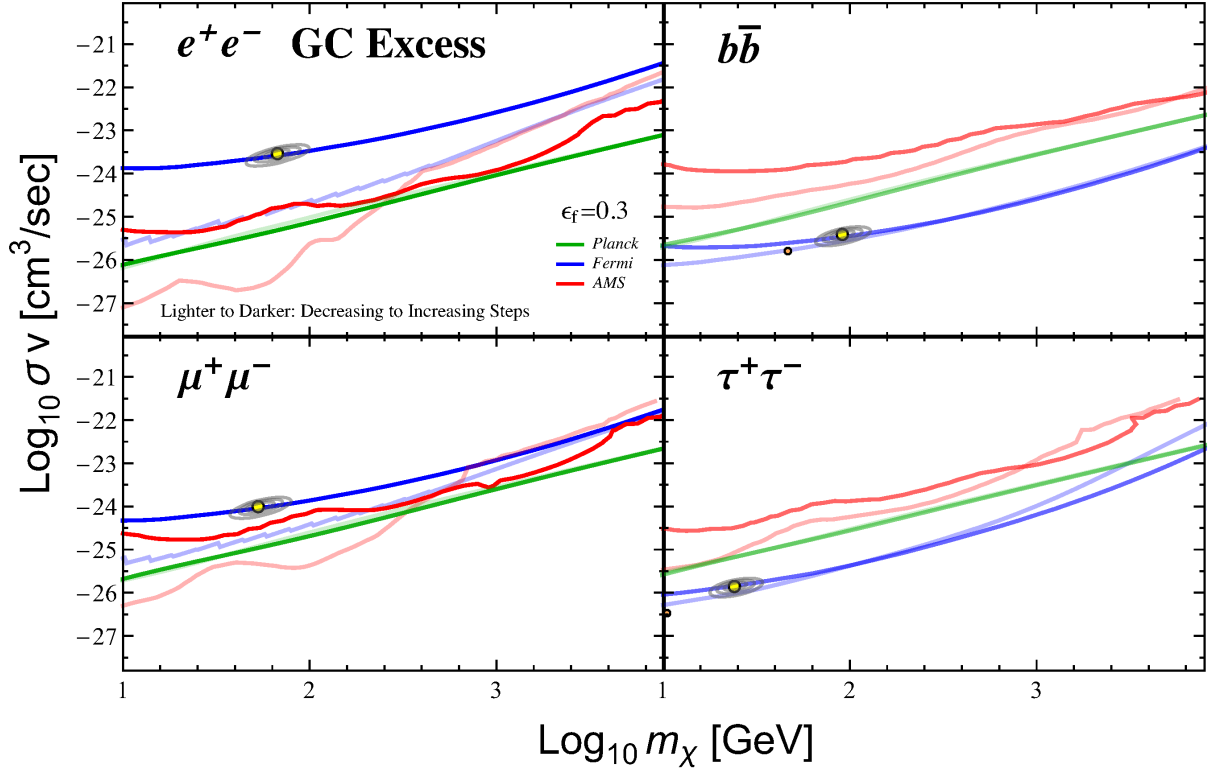}
\end{tabular}
\caption{\footnotesize{The yellow dot corresponds to the overall best fit point to the Galactic Center Excess for a given channel; electrons, muons, taus, and $b$-quarks computed in \cite{Elor:2015bho}. For best fit point we also display the 1,2 and 3 $\sigma$ bands in gray. We have overlaid this with ($95\%$ confidence) constraints on $\langle \sigma v \rangle$ corresponding to the best fit step shown in Table~\ref{table:results0p3}. We also show for each channel the constraints for the case of direct annihilation to DM, and for the $\tau$ and $b$-channels we plot (orange point) the overall best fit for direct annihilation corresponding to $m_\chi \simeq 46.6 \textrm{GeV}$ and $\langle \sigma v \rangle \simeq 1.60 \times 10^{-26} \textrm{cm}^3 \textrm{sec}^{-1}$ for direct annihilations to $b$-quarks and  $m_\chi \simeq 9.96 \textrm{GeV}$ and $\langle \sigma v \rangle \simeq 0.337 \times 10^{-26} \textrm{cm}^3 \textrm{sec}^{-1}$ for direct annihilations to $\tau^+ \tau^-$\cite{Calore:2014xka}.}}
\label{fig:GCE}
\end{figure*}

As an example application, we use our present results to explore the indirect detection constraints on DM explanations for the Galactic Center Gamma-Ray excess observed by \emph{Fermi}, in the context of multi-step cascade annihilations \cite{Elor:2015bho}. Table \ref{table:results0p3} summarizes the results of \cite{Elor:2015bho}: the best-fit values of DM mass, cross section and step number for annihilations to electron, muon, tau and $b$-quark final states (note that other final states considered in the present work will be similar). In Fig.~\ref{fig:GCE} we show these best-fit points for each channel, and the corresponding 1, 2 and 3$\sigma$ confidence contours, together with the indirect detection constraints for that step number and channel. For comparison, we also show the indirect detection constraints for the case of direct annihilation; in this case the best fit occurs for annihilation to $b$-quarks \cite{Calore:2014xka}. (The best-fit points for direct annihilation to $\tau^+ \tau^-$ and $\mu^+\mu^-$ final states occur at $m_\chi \approx 10$ GeV and 5 GeV respectively, with corresponding cross-sections of $3\times 10^{-27}$ cm$^3$/s and $1.6 \times 10^{-26}$ cm$^3$/s.)

In all cases, the best-fit points lie very close to the bound from the \emph{Fermi} dwarf observations. While the constraints can shift as a function of step number, the best-fit cross section and mass shifts in such a way as to maintain this borderline tension. This is not surprising, as \emph{Fermi} observations of the Galactic center and dwarf galaxies probe exactly the same signal, i.e. photons within the energy range of \emph{Fermi}. Any model that reproduces the GeV excess will yield a similar signal in dwarf galaxies.

The \textit{Planck} and AMS-02 bounds, in contrast, do not share this degeneracy. However, for the $e^+ e^-$ and $\mu^+ \mu^-$ states where these constraints dominate, the region of interest for the Galactic Center excess is strongly excluded by both the robust \textit{Planck} limits and the AMS-02 positron bounds.
%

\begin{table}[h]\vspace{0.18in}
\begin{center}
\begin{tabular}{| c || c | c | c | c |}
    \hline
    Final State & $n$-step & $m_\chi$ (GeV)& $\sigma v$ ($\textrm{cm}^3/\textrm{sec}$) & $\chi^2$ \\ \hline \hline
    e & 5 & 67.2 & $2.9 \times 10^{-24}$ & 26.82 \\ \hline
    $\mu$ & 4 & 53.0 & $9.9 \times 10^{-25}$ & 26.94 \\ \hline
    $\tau_{\textrm{unphysical}}$ & 4 & 59.4 & $4.6 \times 10^{-26}$ & 24.13 \\ \hline
    $\tau_\textrm{physical}$ & 2 & 24.1 & $1.4 \times 10^{-26}$ & 25.59 \\ \hline
    $b$ & 2 &  91.2 & $ 3.9 \times 10^{-26}$ & 22.42 \\
    \hline
\end{tabular}
\end{center}
\caption{Best fit to DM annihilations to various final states with $\epsilon_f = 0.3$. For the case of taus we show a best fit point if we include kinematically disallowed masses (unphysical) and also if we restrict ourselves to physical masses as discussed in Sec.~\ref{sec:Review}. Fits were performed over 20 degrees of freedom.}
\label{table:results0p3}
\end{table}

In order for indirect constraints on DM interpretations of the Galactic Center GeV excess to weaken significantly with increased step number, we see that the dominant bound should \emph{not} arise from \emph{Fermi} observations of dwarf galaxies. This could occur, for example, in the case where the final state is a $1:1:1$ admixture of $e^+ e^-$, $\mu^+\mu^-$ and $\tau^+ \tau^-$, with the $\tau^+\tau^-$ component producing the gamma rays required to fit the GeV excess. Such a scenario has been proposed to explain hints of a low-energy counterpart to the excess, arising from inverse Compton scattering and bremsstrahlung of the electrons and positrons \cite{Abazajian:2014hsa}. In the case of direct annihilation, limits from AMS-02 on the $e^+ e^-$ component severely constrain this scenario.

\begin{figure*}[t!]
\centering
\begin{tabular}{c}
\includegraphics[scale=0.7]{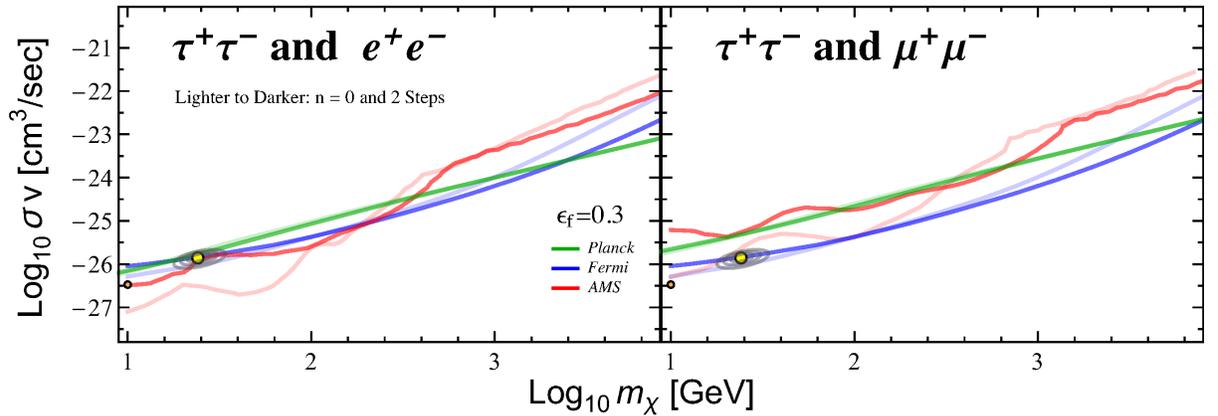}
\end{tabular}
\caption{\footnotesize{Here we consider the scenario in which the final state is an equal mix of electrons, muons and taus, with either direct annihilation or a 2-step cascade (the latter gives the best fit to the GCE for taus). The left panel shows AMS-02 (red) and \textit{Planck} (green) bounds on the electron component; the right panel shows the same bounds on the muon component. Both panels show the dwarf limits on the tau component (blue). Darker lines and the yellow point indicate the bounds and best fit to the GCE, respectively, for the two-step cascade. Fainter lines and the orange point indicate the bounds and best fit to the GCE, respectively, for direct annihilation. The direct-annihilation case appears ruled out by the AMS-02 constraint on the electron component; the two-step cascade is not  excluded by any of these bounds.}}
\label{fig:GCEtau}
\end{figure*}

However, if we instead consider a 2-step cascade, which provides the best fit to the GeV excess for a $\tau^+ \tau^-$ final state, we should instead examine the \textit{Planck} and AMS-02 bounds on 2-step annihilation to muon and electron final states with a DM mass of $\sim 24$ GeV, with a cross section of $\sim 1.4 \times 10^{-26}$ cm$^3$/s. None of the bounds we have derived on the two-step cascade can clearly exclude this scenario, as shown in Figure \ref{fig:GCEtau}.

\clearpage
\newpage
\section{Conclusion}
\label{sec:conclusion}

We have shown that results from current DM indirect searches can be extended to constrain a broad space of dark sector models. We summarize our main points below:
\begin{itemize}
\item{Photon rich final states are generally most constrained by bounds from the {\it Fermi} dwarfs.}
\item{Electron and muon final states are generally most constrained by AMS-02.}
\item{The CMB bounds from {\it Planck}  are robust and insensitive to number of dark sector steps. As a result the CMB bounds may become the most limiting in certain cases where the AMS-02 or dwarf bounds weaken as a result of large $m_\chi$ or increasing number of dark sector steps.}
\item{We find that for a fixed DM mass and final state, the presence of a hidden sector can change the overall cross section constraints by up to an order of magnitude in either direction (although the effect can be much smaller).}
\end{itemize}
For hadronic SM final states ($b$-quarks, gluons, gauge bosons, Higgses), constraints from gamma-ray studies of dwarf galaxies generally remain the most limiting, and -- within the kinematically allowed region -- are generally fairly robust, although they can weaken at low masses and strengthen at high masses. More specifically for small but kinematically allowed masses the bound for final state gauge bosons, Higgses and $b$-quarks can weaken by about 0.1 decades. For the gluon final state, where very low DM masses are in principle possible, this bound can weaken by up to 1.1 decade; however a careful consideration of this regime would require taking into account the mass of the mediators, which may be comparable to $\Lambda_\mathrm{QCD}$. At high masses the bounds will strengthen by about 0.3-0.5 decades for the hadronic final states. 

The photon-rich tau and photon final states behave similarly, with the dwarf limits dominating the constraints except perhaps at very high masses (where it may be important to take constraints from VERITAS and H.E.S.S. into account). Adding extra cascade steps has little effect on the dwarf constraints on the photon final state at low masses (after the addition of the first cascade step, which weakens the limit by up to 0.8 decades), whereas for the tau final state it can weaken the bound by about 0.1 decades within the kinematically allowed regime.

For leptonic final states with few photons (electrons and muons), constraints from AMS-02 often appear to dominate the limits, but are quite sensitive to the number of cascade steps (as well as assumptions on the cosmic-ray propagation and local magnetic field; our limits are more conservative than others in the literature). At low masses (below a few hundred GeV), increasing the number of cascade steps can weaken the constraints by up to 2 orders of magnitude, at which point bounds from the CMB become more constraining. Above a few hundred GeV, however, adding more cascade steps tends to strengthen the constraints, so using results quoted for direct annihilation gives conservative bounds; the CMB limits are also generically stronger than the AMS-02 limits in this mass range. 

If a quick estimate of constraints is needed, the CMB limits almost always appear to be within an order of magnitude of the strongest limit, for the cases we have tested, and vary by at most a factor of 1.3 to 1.5 over cascades with up to 6 steps. 

The details of our code for $n$-step cascades which were used to produced our results are described in App.~\ref{app:fileoutline} and are available at: \\ http://web.mit.edu/lns/research/CascadeSpectra.html.

\section*{Acknowledgements}

We would like to thank Ilias Cholis, Marco Cirelli, Alex Drlica-Wagner, Tim Linden, Lina Necib, Jessie Shelton and Jesse Thaler for helpful discussions and comments. We would like to further thank Lina Necib for carefully testing our publicly released code, and Rich Pieri for assistance with the code release. We thank the anonymous referee for very helpful comments, which improved the manuscript. This work is supported by the U.S. Department of Energy under grant Contract Numbers DE$-$SC00012567 and DE$-$SC0013999. NLR is supported by the Acevedo Fellowship.

\begin{figure*}[htbp]
\hspace*{-1cm}
\begin{tabular}{c}
\includegraphics[scale=0.8]{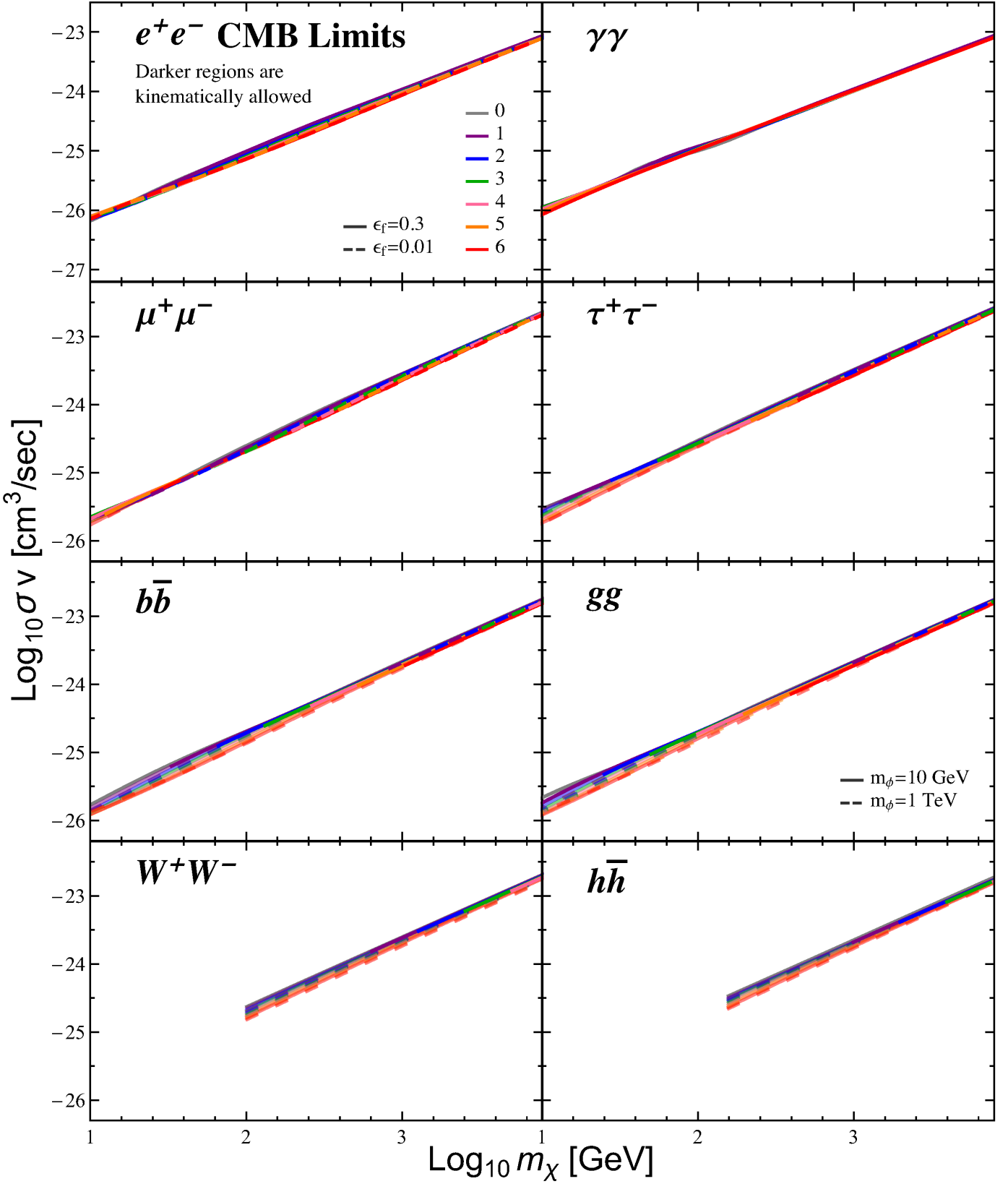} \hspace{0.1in}
\end{tabular}
\caption{\footnotesize{The $95\%$ confidence bound on DM annihilation cross-section (Eq.~\ref{eq:CMBbound})  for $n=$ 1-6 step cascade for various final states, with $\epsilon_f = 0.3$ (solid) and  $\epsilon_f = 0.01$ (dashed). The shaded out portions of the plot correspond to values of $m_\chi$ that are kinematically forbidden. As discussed above the number of steps does not affect the total power deposited by the DM annihilation per unit time. Therefore the constraints are insensitive to the number of steps as the only impact comes from the energy dependence of $f_{\textrm{eff}}$ and the broadening of the spectrum.}}
\label{fig:CMB0p3}
\end{figure*}

\begin{figure*}[htbp]
\hspace*{-1.1cm}
\includegraphics[scale=0.8]{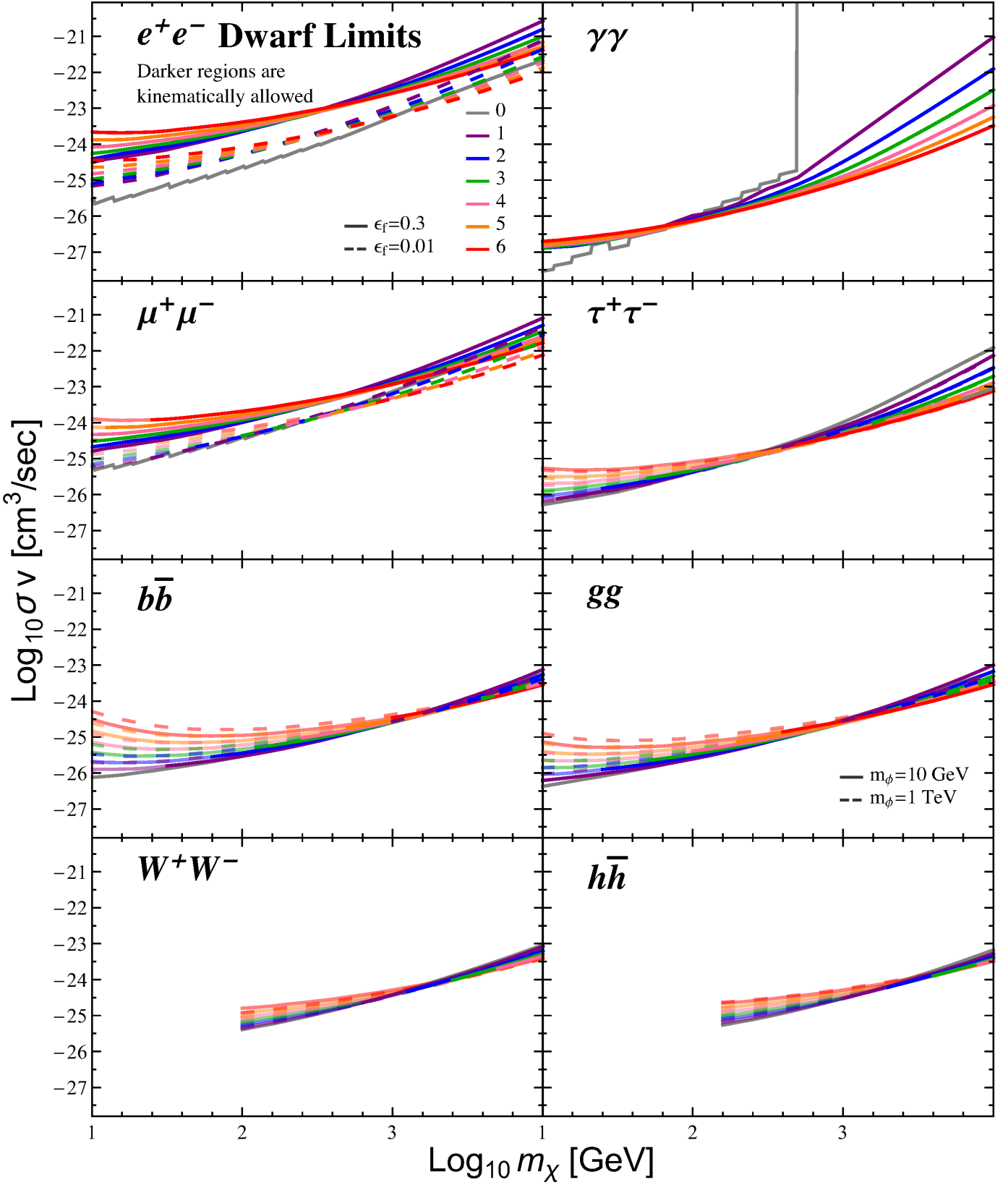}
\caption{\footnotesize{95\% confidence limits on DM cross-section for cascade models using the data from 15 dwarf spheroidal galaxies. Results are shown for the photon spectrum obtained from eight different final states: electrons, photons, muons, taus, $b$-quarks, gluons, $W$-bosons, and Higgs. In each case we show the results of a 0 (direct), or 1-6 step cascade. Additionally where it makes sense we show results for two different $\epsilon_f$ values, solid lines representing $0.3$  and dashed $0.01$. Note the $\gamma \gamma$ spectrum is independent of $\epsilon_f$, so we only show one set of limits there, and for the gluon spectrum the relevant parameter is instead $m_{\phi}$ and we show results for $10$ GeV in solid and $1$ TeV as dashed. Only the darker regions are kinematically allowed. See text for a discussion of the results.}}
\label{fig:DwarfLimits}
\end{figure*}

\clearpage

\appendix
\section{Details of $n$-body Cascades}
\label{app:multibody}

\begin{figure*}[t!]
\centering
\begin{minipage}{.45\textwidth}
  \centering
  \includegraphics[scale=0.7]{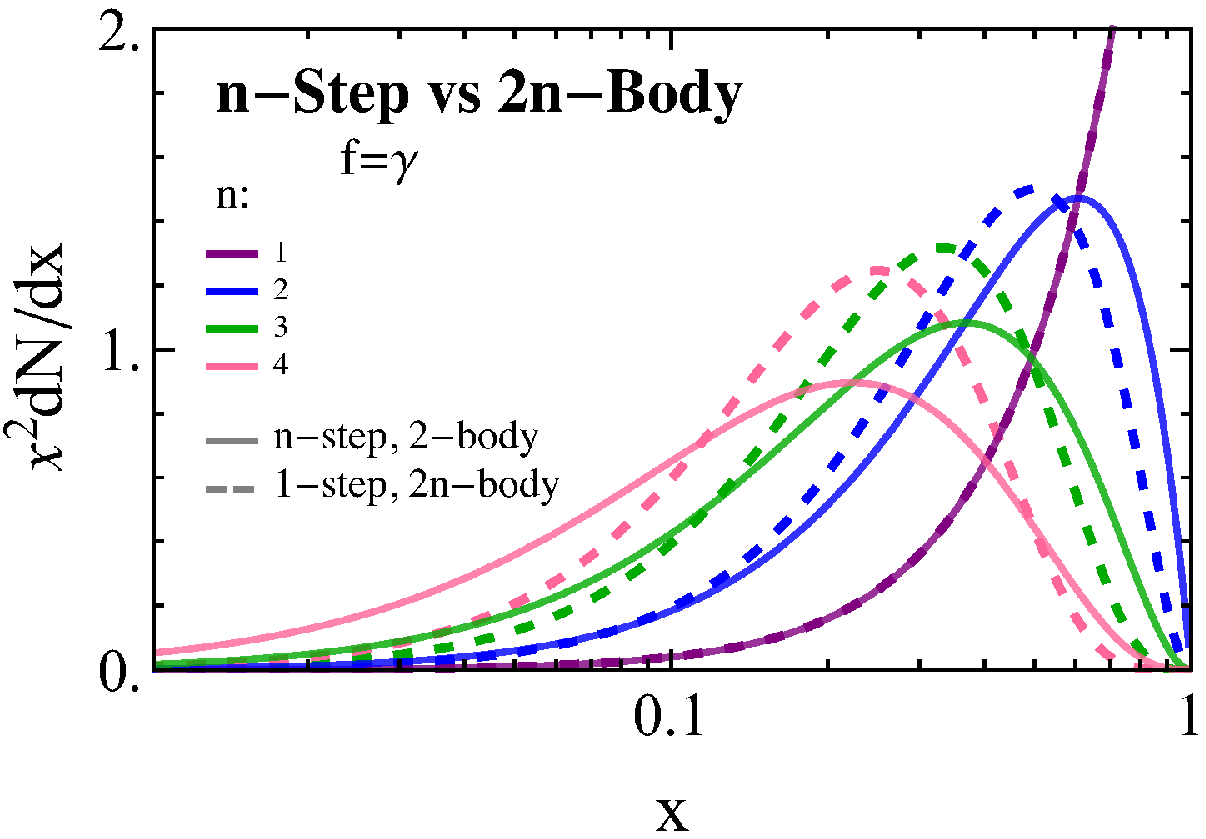}
  \captionof{figure}{\footnotesize{Spectra for a cascades containing $n$-step 2-body decay and a 1-step $2n$-body decay, both to $\gamma \gamma$, are shown as the solid and dashed curves respectively, for the case of $n=$ (1,2,3,4) in (purple, blue, green, pink). We see that for $n=$ 1 and 2 we can approximate one of these types of spectra by the other (with the $n=$ 1 case being exact by definition).}}
  \label{fig:nbodyvsnstepgamma}
\end{minipage}
\hspace{0.4in}
\begin{minipage}{.45\textwidth}
\vspace{-0.4in}
  \centering
  \includegraphics[scale=0.7]{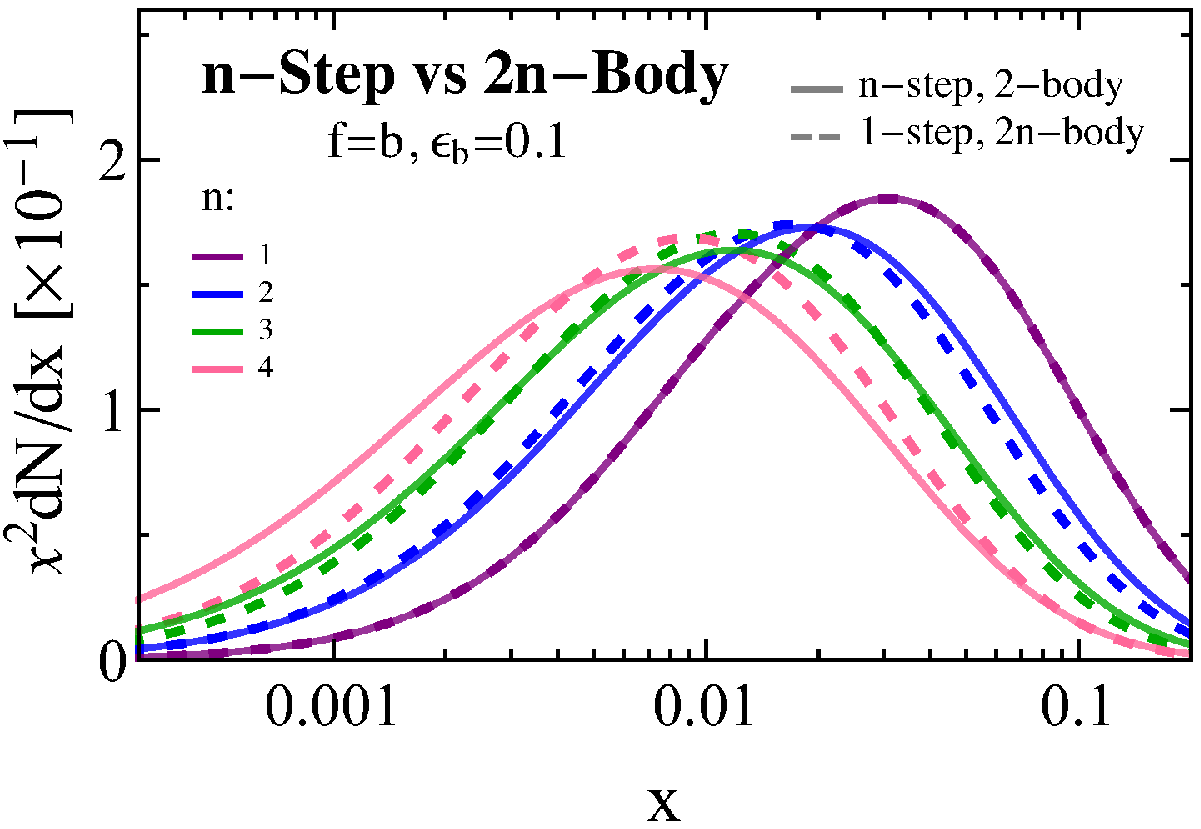}
  \captionof{figure}{\footnotesize{The same as Fig.~\ref{fig:nbodyvsnstepgamma}, but for a final state $b\bar{b}$ with $\epsilon_b=0.1$. Note that again we get close agreement in the $n=$ 1 and 2 case.}}
  \label{fig:nbodyvsnstep}
\end{minipage}
\end{figure*}

\begin{figure}[t!]
\centering
  \includegraphics[scale=0.7]{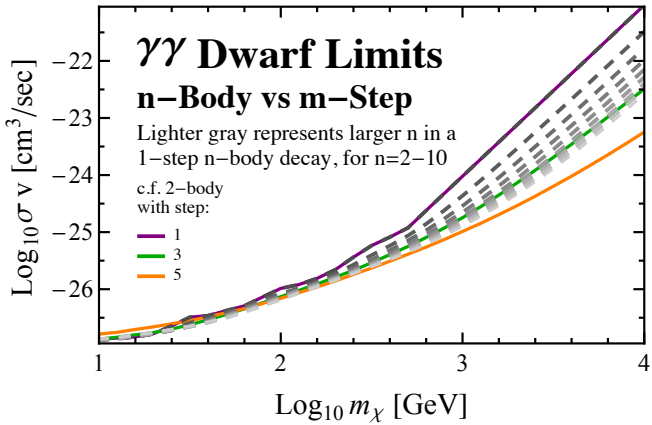}
  \caption{\footnotesize{Dwarf limits for n-body vs m-step cascades for the $\gamma \gamma$ final state. We show the multi-body case for $n=$ 2-10 in gray, with lighter gray corresponding to larger $n$. In orange, green and purple we also show the 1, 3 and 5-step 2-body cascade for the same final state. As discussed in the text, for the multi-body case the spectrum sits in between the cascade spectra, and thus we expect the limits to do the same. The figure makes this clear and emphasises how the multi-body framework is captured within the cascade setup.}}
  \label{fig:DwarfdeltaNBody}
\end{figure}

In this appendix we will derive Eq.~\ref{eq:nbodyboosteq} and provide some additional intuition for this case as well as pointing out that for a small number of steps, the cascade setup can provide an excellent approximation (albeit with some dependence on the channel). To set up this problem, firstly recall that the key physics encapsulated in Eq.~\ref{eq:boosteq} is that when we add in a cascade step we need to boost the spectrum to the new rest frame. In the case of 2-body decays this is particularly simple, because we know exactly how much to boost by. Explicitly, if we have added in a step of the form $\phi_i \to \phi_{i-1} \phi_{i-1}$, then in the $\phi_i$ rest frame we know the $\phi_{i-1}$ particles must be emitted back to back, meaning we know their energy and hence their boost. If instead we introduce a step via $\phi_i \to \phi_{i-1} \phi_{i-1} \phi_{i-1}$, we no longer know the boost exactly, instead we can only associate a probability with any boost which we can determine from the energy distribution for a given $\phi_{i-1}$. Accordingly what we need to calculate is the energy spectrum of a particular $\phi$ in the decay $\chi \chi \to n \times \phi$, and then combine this with a version of Eq.~\ref{eq:boosteq} suitable for a general boost. Below we will firstly do this exactly for the case of a 3-body decay, show what this becomes after applying the large hierarchies approximation, and then we will show the general $n$-body result assuming hierarchical decays.

As discussed, our starting point is the energy spectrum of a particular $\phi$ in the decay $\chi \chi \to 3 \times \phi$, which can be determined from the three body phase space. For this purpose we make use of the analytic formula for the $n$-body phase space outlined in \cite{Byckling:1969sx,Kersevan:2004yh}. In the case where our three final state scalars have mass $m$, we can write the 3-body phase space as:
\begin{equation}\begin{aligned}
\Phi_3 = (4\pi)^2 \int_{4m^2}^{(M_3-m)^2} &dM_2^2 \frac{\sqrt{\lambda(M_3^2,M_2^2,m^2)}}{8M_3^2} \\
\times &\frac{\sqrt{\lambda(M_2^2,m^2,m^2)}}{8M_2^2}\,,
\label{eq:3bodyphasespace}
\end{aligned}\end{equation}
where $\lambda(x,y,z)=x^2 + y^2 + z^2 - 2xy - 2yz - 2zx$ and if we say the mass of the DM is $m_{\chi}$ and the energy of one $\phi$ particle is $E$, then $M_3^2=4m_{\chi}^2$ and $M_2^2 = 4m_{\chi}^2 + m^2 - 4 m_{\chi} E$. Using this, the energy spectrum of the scalars is simply:
\begin{equation}
\frac{dN_{\phi}}{dE} \propto \frac{d\Phi_3}{dE}\,,
\label{eq:3bodytospec}
\end{equation}
where the constant of proportionality can be determined by normalising the spectrum. Before proceeding, it is useful to introduce a set of dimensionless variables to work with as we did in the 2-body case. As there, we firstly define $\epsilon_1= m/m_{\chi}$, but note here that $\epsilon_1 \in [0,2/3]$, rather than $[0,1]$ as in the 2-body case. To play a similar role to $x$, we also introduce $\xi = E/m_{\chi} \in [\epsilon_1,1-3\epsilon_1^2/4]$, where the limits here are fixed by Eq.~\ref{eq:3bodyphasespace} and can also be seen from the kinematics. In terms of these variables, we can use Eq.~\ref{eq:3bodytospec} and Eq.~\ref{eq:3bodyphasespace} to arrive at:
\begin{equation}
\frac{dN_{\phi}}{d\xi} = C \sqrt{\frac{(\xi^2-\epsilon_1^2)(4-3\epsilon_1^2-4\xi)}{4+\epsilon_1^2-4\xi}}\,,
\label{eq:3bodyspec}
\end{equation}
where $C$ is a constant that normalises the spectrum and can be determined numerically. Note that when $\epsilon_1 \to 2/3$, this distribution approaches a $\delta$ function, as expected when the particles are all produced at rest. We will return to the limit of small $\epsilon_1$ shortly.

Using this result, we can then revisit the derivation of the boost formula given in \cite{Elor:2015tva}, a hierarchical version of which is given in Eq.~\ref{eq:boosteq}, and derive the analogue for an arbitrary boost. Doing so, if we label the spectrum of the decay of $\phi \to 2 \times \text{(SM final state)}$ as $dN/dx_0$, we can write the spectrum of the same particle from the decay $\chi \chi \to 3 \times \phi \to 6 \times \text{(SM final state)}$ as:
\begin{equation}\begin{aligned}
\frac{dN}{dx_1} = 3 &\int_{\epsilon_1}^{1-(3/4)\epsilon_1^2} d \xi C \sqrt{\frac{(\xi^2-\epsilon_1^2)(4-3\epsilon_1^2-4\xi)}{4+\epsilon_1^2-4\xi}} \\
\times &\int_{t_{\rm min}}^{t_{\rm max}} \frac{dx_0}{x_0 \sqrt{\xi^2-\epsilon_1^2}} \frac{dN}{dx_0}\,,
\label{eq:3bodyspecfull}
\end{aligned}\end{equation}
where we have defined:
\begin{equation}\begin{aligned}
t_{\rm max} &\equiv {\rm min} \left[ 1, \frac{2x_1}{\epsilon_1^2} \left( \xi + \sqrt{\xi^2-\epsilon_1^2} \right) \right] \\
t_{\rm min} &\equiv \frac{2x_1}{\epsilon_1^2} \left( \xi - \sqrt{\xi^2 - \epsilon_1^2} \right)
\end{aligned}\end{equation}
There are two directions the above result can be generalised. For one, we could extend this to a longer cascade of 3-body decays, although the logic here is identical to the general 2-body case discussed in \cite{Elor:2015tva}, so we will not repeat that here. Secondly we can look to extend this to 4-body decays and higher. The difficulty with this is that the $n$-body phase space quickly becomes analytically intractable. Nevertheless as observed in \cite{Liu:2014cma}, in the large hierarchies regime ($\epsilon_1 \ll 1$) we regain analytic control as we will now outline.

\begin{figure*}[t!]
\centering
\begin{minipage}{.45\textwidth}
  \centering
  \includegraphics[scale=0.65]{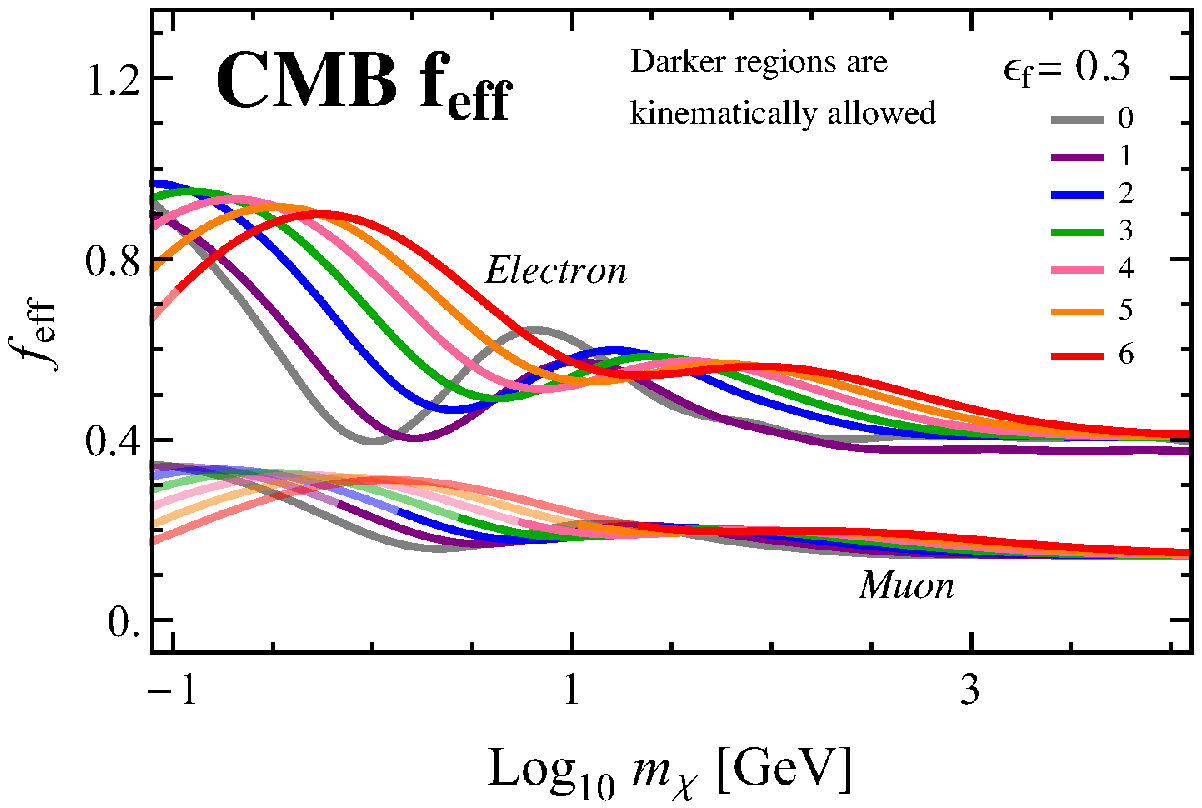}
  \captionof{figure}{\footnotesize{$f_\mathrm{eff}$ for $n =$ 0-6 step cascades to final state electrons and muons, with $\epsilon_f = 0.3$. Note that the difference in pattern between $f_\mathrm{eff}$ for direct and single step electrons and higher step cascades can be understood by recalling that the direct electron FSR spectrum is sharply peaked. Each subsequent cascade step smooths out this spectrum thus changing the shape significantly.}}
  \label{fig:feffElectronMuon}
\end{minipage}
\hspace{0.4in}
\begin{minipage}{.45\textwidth}
\vspace{-0.65in}
  \centering
  \includegraphics[scale=0.665]{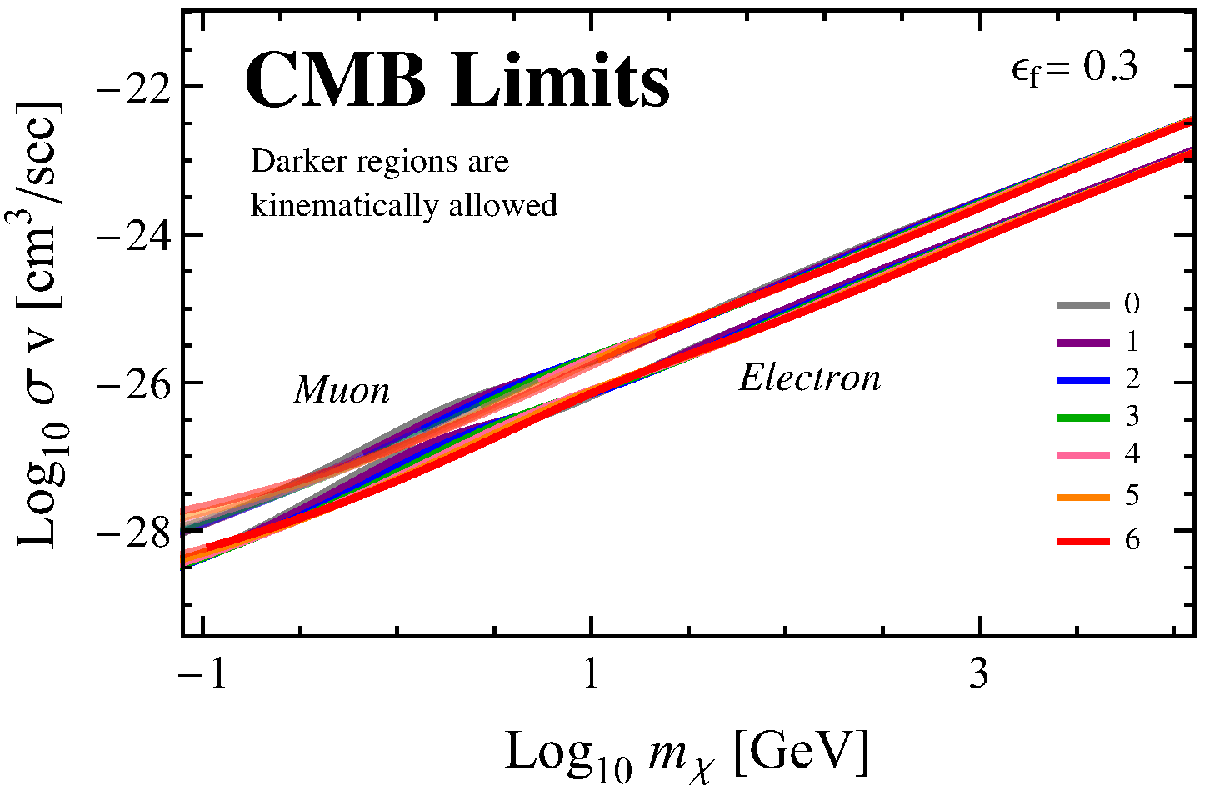}
  \captionof{figure}{\footnotesize{The bound on $\langle \sigma v \rangle$ for $n=$ 0-6 step cascades to final state electrons and muons, with $\epsilon_f = 0.3$.}}
  \label{fig:BoundElectronMuon}
\end{minipage}
\end{figure*}

Returning to Eq.~\ref{eq:3bodyspec}, taking the $\epsilon_1 \to 0$ limit we find that:
\begin{equation}
\frac{dN_{\phi}}{d\xi} = 2 \xi + \mathcal{O}(\epsilon_1^2)\,,
\label{eq:3bodysmalleps}
\end{equation}
where now $\xi \in [0,1]$. Following \cite{Liu:2014cma}, this can then be generalised to the $n$-body case, where we find:
\begin{equation}
\frac{dN_{\phi}}{d\xi} = (n-1)(n-2)(1-\xi)^{n-3} \xi + \mathcal{O}(\epsilon_1^2)\,,
\label{eq:nbodysmalleps}
\end{equation}
where again $\xi \in [0,1]$. Using this we can finally give the equivalent expression of Eq.~\ref{eq:boosteq} for the $n$-body case:
\begin{equation}
\frac{dN}{dx_1} = n(n-1)(n-2) \int_0^1 d \xi (1-\xi)^{n-3} \int_{x_1/\xi}^1 \frac{dx_0}{x_0} \frac{dN}{dx_0} + \mathcal{O}(\epsilon_1^2)
\label{eq:nbodyboosteqapp}
\end{equation}
thereby demonstrating Eq.~\ref{eq:nbodyboosteq}.

As a simple example of how this can be used, consider the decay $\phi \to \gamma \gamma$ which has the spectrum $dN_{\gamma}/dx_0 = 2 \delta(x_0-1)$. If we substitute this in, we find the spectrum for $\chi \chi \to n \times \phi \to 2n \times \gamma$ is just
\begin{equation}
\frac{dN_{\gamma}}{dx_1} = 2n(n-1)(1-x_1)^{n-2}\,.
\label{eq:nbodydelta}
\end{equation}
Integrating this over $x_1 \in [0,1]$, we find $N_{\gamma}=2n$, as expected.

To follow on from this, consider the spectrum derived by repeated application of the boost formula in Eq.~\ref{eq:boosteq} to the same $\phi \to \gamma \gamma$ spectrum, $dN_{\gamma}/dx_0 = 2 \delta(x_0-1)$. Doing so we obtain:
\begin{equation}
\frac{dN_{\gamma}}{dx_n} = \frac{(-2)^{n+1}}{(n-1)!} \ln^{n-1} x_n\,.
\label{eq:nstepdelta}
\end{equation}
Note that if we integrate this over $x_n \in [0,1]$, we find $N_{\gamma}=2^{n+1}$. Now by definition Eq.~\ref{eq:nbodydelta} with $n=$ 2 is identical Eq.~\ref{eq:nstepdelta} with $n=$ 1, as in this case they both represent a 2-body 1-step cascade. Note also though that if we take Eq.~\ref{eq:nbodydelta} with $n=$ 4 and Eq.~\ref{eq:nstepdelta} with $n=$ 2, then both situations have the same number of final state photons from different kinematic setups. In Fig.~\ref{fig:nbodyvsnstepgamma} we compare an $n$-step 2-body decay and a 1-step $2n$-body decay for final state photons, for $n=$ (1,2,3,4). We see that whilst they agree for $n=$ 1 (by definition), and are quite similar to each other for $n=$ 2, this similarity breaks down rapidly. This is not entirely surprising as a 6-body 1-step cascade has a different number of final state photons to a 2-body 3-step cascade, but one can also check that this latter spectrum also does not agree well with an 8-body 1-step result which would have the same number of photons. In Fig.~\ref{fig:nbodyvsnstep} we show the same comparison for final state $b\bar{b}$ with $\epsilon_b=0.1$. Here we see the agreement is better although still beginning to break down for larger $n$. We also tested this for several other final states, with the common theme that the spectrum of a 4-body 1-step decay is often well approximated by that of a 2-step 2-body decay.

Lastly let us confirm the claim from the main text that the results for multi-body decays sit in between our multi-step cascade results. For this purpose consider again the photon spectrum obtained from $\phi$ decaying into $b\bar{b}$ with $\epsilon_b=0.1$. In this case we plot the $n$-body spectrum for $n=$ 2 to 10 in Fig.~\ref{fig:nbody}, which we presented in the main text. In the figure we also plot the case of a 1-step, 3-step and 5-step 2-body cascade. Clearly the $n$-body results sit in between these multi step cascade cases, which indicates that the constraints on an $n$-body 1-step cascade will be largely contained within the limits on a 2-body $n$-step cascade. To demonstrate this, in Fig.~\ref{fig:DwarfdeltaNBody} we show that for the case of the $\gamma \gamma$ limits extracted from the dwarfs, the $n=$ 2-10 multi-body decays sit exactly in between the 1, 3 and 5-step cascades as claimed.

\section{Details of the CMB Results}
\label{app:CMBdetails}

In this appendix we present additional results from the CMB analysis. For many of the final states considered in the main text, the kinematic threshold on their production means that it is not sensible to go to lower masses than we presented. This is not the case, however, for electrons and muons, where we can take our results to much lower masses. 

\begin{figure*}[htbp]
\hspace*{-0.5cm}
\includegraphics[scale=0.78]{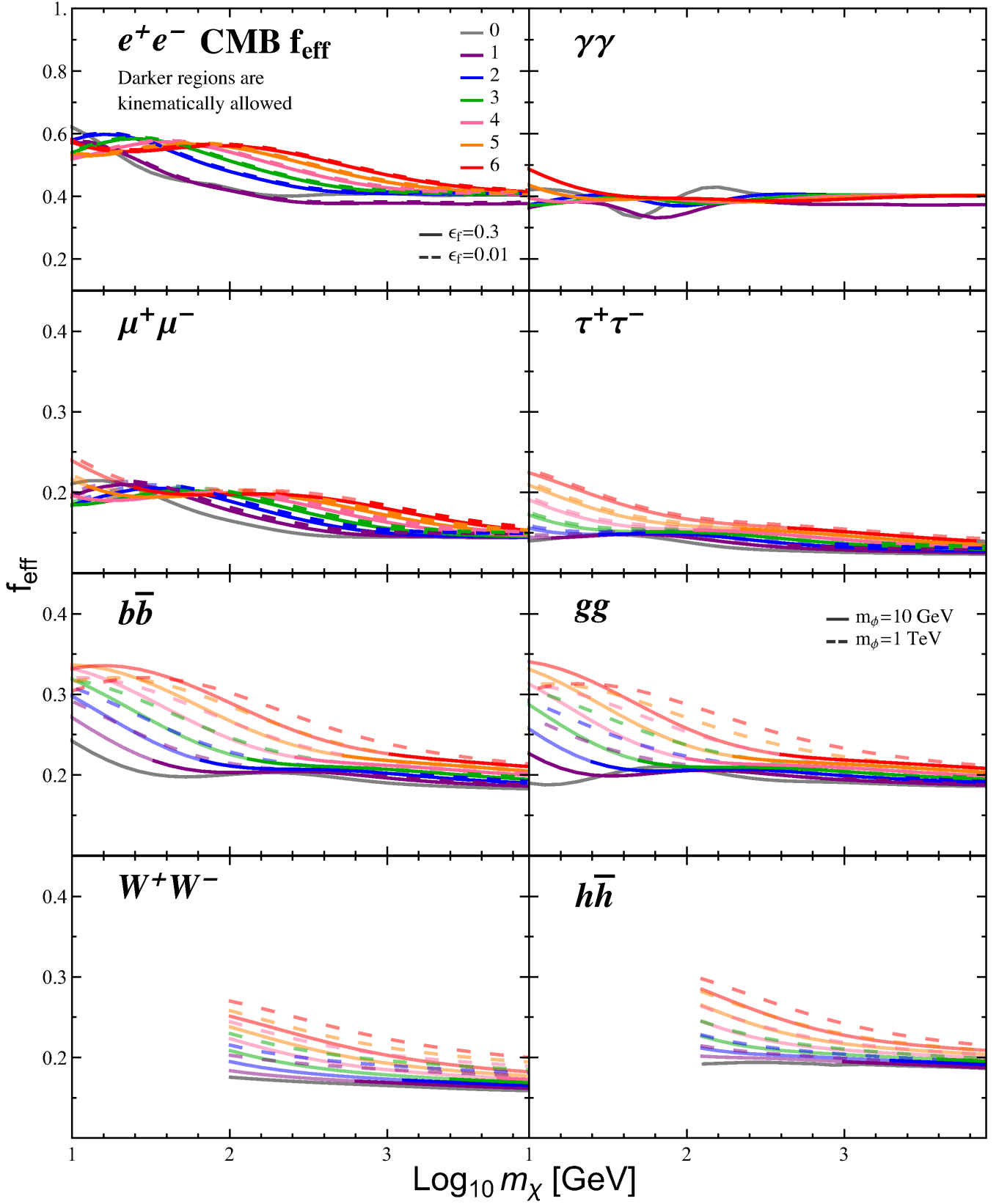}
\caption{\footnotesize{$f_\mathrm{eff}$ for $n=$ 1-6 step cascade for various final states, with $\epsilon_f = 0.3$ (solid) and  $\epsilon_f = 0.01$ (dashed). The shaded out portions of the plot correspond to values of $m_\chi$ that are kinematically forbidden. For the case of direct annihilation (gray line) only the spectrum for $m_\chi >10$GeV is displayed, since for lower values of $m_\chi$ the PPPC is unreliable. For direct annihilations to photons the spectrum is simply a delta function so in this case we plot $f_\mathrm{eff}$ down to lower masses as well.}}
\label{fig:feffCMB0p3}
\end{figure*}

In Fig.~\ref{fig:feffElectronMuon} we present the value of $f_{\textrm{eff}}$ for cascades ending in final state electrons and muons with $\epsilon_f = 0.3$. Here we consider DM with mass as low $\mathcal{O}({\rm keV})$ which is relevant to various CMB studies, which should be compared with our general results for $f_\textrm{eff}$ in Fig.~\ref{fig:feffCMB0p3} for DM annihilations into the eight final states considered in the main text. As expected $f_{\textrm{eff}}$ is largest for annihilations to final state electrons.

\begin{figure*}[htbp]
\hspace*{-1.45cm}
\vspace*{-0.15cm}
\includegraphics[scale=0.82]{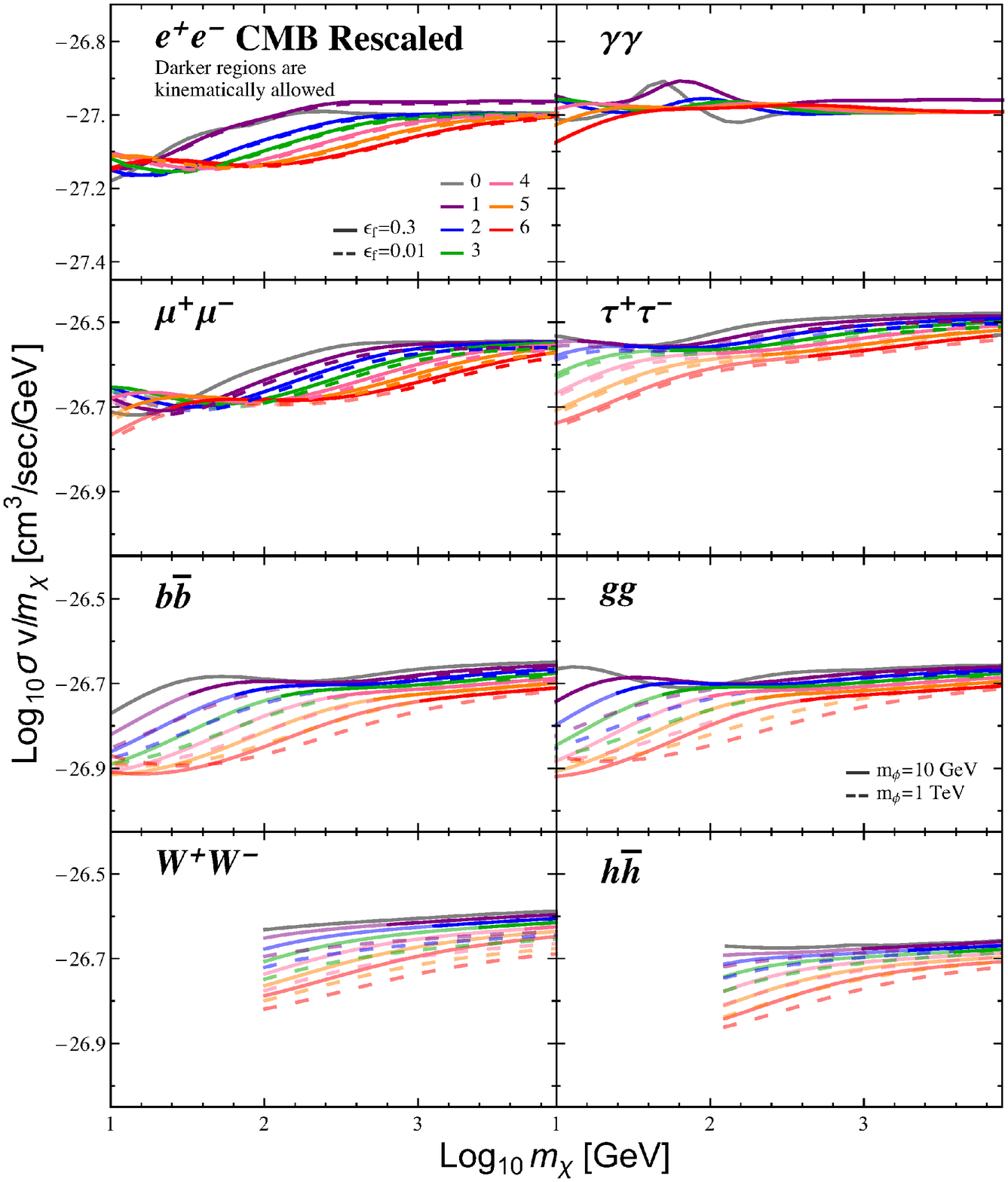}
\caption{\footnotesize{Values of the bound on $\langle \sigma v \rangle / m_\chi$ for various final states. The bound is very robust; we find roughly $\langle \sigma v \rangle/m_\chi \lesssim 10^{-27.3} - 10^{-26.6}~\textrm{cm}^3/\textrm{s}/\textrm{GeV}$, independent of final state (although the bound is slightly higher for electrons and photons), number of steps, or $\epsilon_f$.
}}
\label{fig:ScaledCMB0p3}
\end{figure*}

The corresponding bound $\langle \sigma v \rangle$ for a given $m_{\chi}$ for light DM annihilating to final state electrons and muons is displayed in Fig ~\ref{fig:BoundElectronMuon}, and more generally in Fig ~ \ref{fig:CMB0p3}. As this bound is fairly insensitive to the final state and number of steps it is interesting to examine the rescaled bound on $\langle \sigma v \rangle / m_\chi$ which we display in Fig.~\ref{fig:ScaledCMB0p3} for $n =$ 0-6 step cascades to various final states. We find that generically the re-scaled bound for all SM final states, $\epsilon_f$ values, and cascade step falls within the narrow range $\langle \sigma v \rangle / m_\chi = 10^{-27.3} -  10^{-26.6}~\textrm{cm}^3/\textrm{s}/\textrm{GeV}$.

\section{Pass 7 versus Pass 8 for the Dwarfs}
\label{app:p7p8}

\begin{figure*}[t!]
\centering
\begin{tabular}{c}
\includegraphics[scale=0.7]{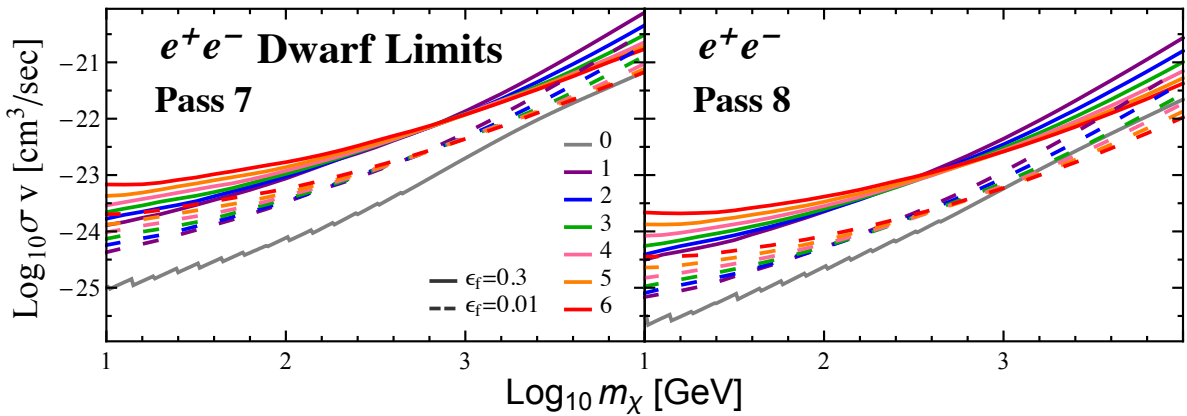}
\end{tabular}
\caption{\footnotesize{Here we recreate the results shown in Fig.~\ref{fig:DwarfLimits} for the case of final state $e^+ e^-$, for the case of the 4 years of Pass 7 data analysed in \cite{Ackermann:2013yva} (left) and for the 6 years of Pass 8 data considered in \cite{Ackermann:2015zua} (right). We see that the updated dataset essentially just strengthens the limits by roughly half an order of magnitude, without noticeably changing other basic features.}}
\label{fig:P7vsP8Dwarfs}
\end{figure*}

As discussed in the main body, the limits displayed in Fig.~\ref{fig:DwarfLimits} were derived using 6 years of Pass 8 data collected using the {\it Fermi} Gamma-Ray Space Telescope; more specifically using the publicly available results of \cite{Ackermann:2015zua} made from analysing this data. This work was an updated version of the analysis that appeared in \cite{Ackermann:2013yva}, which set limits using the same 15 dwarf spheroidal galaxies, but only with 4 years of Pass 7 data. These results are also publicly available,\footnote{Both can be obtained from \\ http://www-glast.stanford.edu/pub\_data/} meaning we can cross check how much our results change when between datasets. We did this for each of the final states considered in Fig.~\ref{fig:DwarfLimits} and found generically the shape of the limit curves were unchanged, but that the limits themselves improved by roughly half an order of magnitude when using the updated analysis. We show an example of this for the case of electrons in Fig.~\ref{fig:P7vsP8Dwarfs}, and we see that the generic features of the limits are unchanged but the results strengthen as we move from the Pass 7 to the Pass 8 dataset.

\section{Description of Cascade Spectra Files}
\label{app:fileoutline}

All of the spectra used in this work are publicly available in .dat format at: \\ http://web.mit.edu/lns/research/CascadeSpectra.html. \\ The details of how these spectra were generated from the direct spectra mentioned in Sec.~\ref{sec:Spec} is discussed in Sec.~\ref{sec:Review} and more comprehensively in \cite{Elor:2015tva}. The format of the spectra files has been modeled after those made available by \cite{Cirelli:2010xx}, in the hope that anyone who has used the results of that paper should have no difficulty using ours. In addition to the files themselves we have also included two example files showing how to load the spectra in Mathematica and Python.

There are four basic file types included, which we describe briefly in turn.
\begin{itemize}
\item AtProduction\_\{gammas,positrons,antiprotons\}.dat: these are the files provided by \cite{Cirelli:2010xx} and contain the 0-step or direct annihilation spectrum of \{photons, positrons, antiprotons\} for various final states;
\item Cascade\_\{Gam,E,Mu,Tau,B,W,H,G\}\_gammas.dat: photon spectrum from final state \{photons, electrons, muons, taus, $b$-quarks, Ws, Higgs, gluons\};
\item Cascade\_\{Gam,E,Mu,Tau,B,W,H,G\}\_positrons.dat - positron spectrum from final state \{photons, electrons, muons, taus, $b$-quarks, $W$s, Higgs, gluons\}; and
\item Cascade\_\{B,W,H,G\}\_antiprotons.dat - antiproton spectrum from final state \{$b$-quarks, $W$s, Higgs, gluons\}.
\end{itemize}
Again we emphasise that the AtProduction files were created by the authors of \cite{Cirelli:2010xx}, we only include them in our results as it is convenient to store the 0-step spectra in separate files from the cascade results, yet having them in the same place is useful.

As for the contents of the files, firstly the three AtProduction\_\{gammas,positrons,antiprotons\}.dat have the following format:
\begin{itemize}
\item Each file has 30 columns and 11099 rows, where the first row contains column labels and all others contain numerical values.
\item The first column contains the DM mass in GeV, running from 5 GeV up to 100 TeV. Note using these direct spectra below 5 GeV is not advised as the extrapolation is often unreliable.
\item The second column contains $\log_{10}(x)$ values, where $x=E/m_{\chi}$. This ranges from -8.9 to 0 in steps of 0.05.
\item Finally the columns 3-30 contain the value of the spectrum in $dN/d\log_{10}(x) = \ln(10) x dN/dx$ of the spectrum at that value of $m_{\chi}$ and $x$. The columns of relevance for us are 5 (electrons), 8 (muons), 11 (taus), 14 ($b$-quarks), 18 ($W$-bosons), 22 (gluons), 23 (photons) and 24 (Higgs).
\end{itemize}

The contents of the 19 Cascade\_\{Final State\}\_\{Spectrum Type\}.dat has been modeled on these files. To be explicit we have:
\begin{itemize}
\item Each file has 8 columns and 1612 rows, where the first row contains column labels and all others contain numerical values.
\item The first column contains the value of $\epsilon_f$. We include the spectra for the values 0.01, 0.03, 0.05, 0.07, 0.1, 0.2, 0.3, 0.4 and 0.5. The only exception to this is for gluons or the positron spectrum from photons, where the first column contains $m_{\phi}$ values instead, and we include values of 10, 20, 40, 50, 80, 100, 500, 1000 and 2000 GeV. Within these parameter ranges the interpolation is quite reliable, but outside these ranges linear interpolation is recommended. Note that several spectra, such as the $\gamma \gamma$ photons spectrum or the electron positron spectrum have no dependence on $\epsilon_f$ or $m_{\phi}$. Nevertheless we still include an $\epsilon_f$ column in those files for consistency, and note picking any value of this parameter will result in an identical spectrum.
\item The second column contains $\log_{10}(x)$ values, where $x=E/m_{\chi}$. This ranges from -8.9 to 0 in steps of 0.05.
\item Finally the columns 3-8 contain the value of the spectrum in $dN/d\log_{10}(x) = \ln(10) x dN/dx$ of the spectrum at that value of $m_{\chi}$ and $\epsilon_f$ or $m_{\phi}$. The columns represent an $n=$ 1 cascade (column 3) up to an $n=$ 6 one (column 8).
\end{itemize}

\bibliography{cascades}

\end{document}